\documentclass[final]{elsarticle}
\usepackage{lineno,hyperref}
\usepackage{algorithm}
\usepackage{algpseudocode}
\usepackage{amsmath}
\usepackage{amsthm}
\usepackage{amsfonts}
\usepackage{graphicx}
\usepackage{epstopdf}
\usepackage{color}
\usepackage[normalem]{ulem}
\theoremstyle{remark}
\newcommand{\argmin}{\operatornamewithlimits{arg\;min}}
\newcommand{\argmax}{\operatornamewithlimits{arg\;max}}

\newcommand\T{\rule{0pt}{2.6ex}} % Top strut
\newcommand\B{\rule[-1.2ex]{0pt}{0pt}}
\journal{}
\newtheorem{proposition}{Proposition}

\begin{document}
	
	\begin{frontmatter}
		
\title{
%Batch Self Organizing maps for distributional data \textcolor{red}{with computation of weights of relevance for the variables} 
Batch Self Organizing maps for distributional data with automatic weighting of variables and components}
		
%        \author[SUN]{Antonio Irpino and Rosanna Verde}
%        \ead{antonio.irpino@unina2.it, rosanna.verde@unina2.it}
%        \address[SUN]{Second University of Naples, Dept. of Political Sciences, 81100 Caserta, Italy}
%        \author[CIn]{Francisco de A.T. de Carvalho\corref{cor1}}
%\ead{fatc@cin.ufpe.br}
%\cortext[cor1]{Corresponding Author. tel.:+55-81-21268430; fax:+55-81-21268438}
%        \address[CIn]{Centro de Inform\'atica, Universidade Federal de
%Pernambuco, Av. Jornalista Anibal Fernandes, s/n - Cidade Universit\'aria, CEP 50740-560, Recife (PE), Brazil}

		\author{Francisco de A.T. De Carvalho\fnref{myfootnote2}}
		\address{Centro de Informatica, Universidade Federal de Pernambuco,
			Av. Jornalista Anibal Fernandes s/n - Cidade Universitaria,
			CEP 50740-560, Recife-PE, Brazil}
		\fntext[myfootnote2]{Corresponding Author: fatc@cin.ufpe.br}
		\author{Antonio Irpino, Rosanna Verde and Antonio Balzanella \fnref{myfootnote}}
		\address{Universit\'a degli Studi della Campania ``Luigi Vanvitelli'', Dept. of Mathematics and Physics, Viale Lincoln, 5, 81100 Caserta, Italy}
		\fntext[myfootnote]{\{antonio.irpino\},\{rosanna.verde\},\{antonio.balzanella\}@unicampania.it}

		
		\begin{abstract}
This paper deals with Self Organizing Maps (SOM) for data described by distributional-valued variables. 
This kind of variables takes as values empirical distributions on the real line or estimates of probability distributions.
We propose a Batch SOM strategy (DBSOM) that optimizes an objective function, using a $L_2$ Wasserstein distance that is a suitable dissimilarity measure to compare distributional data, already proposed in different distributional data analysis methods.
Moreover, aiming to take into consideration the different contribution of the variables, we propose an adaptive version of the DBSOM algorithm.
This adaptive version has an additional step that learns automatically a relevance weight for each distributional-valued variable.
Besides, since the $L_2$ Wasserstein distance allows a decomposition into two components: one related to the means and one related to the size and shape of the distributions, also relevance weights are automatically learned for each of the measurement components to emphasize the importance of the different estimated parameters of the distributions.
Experiments with real datasets of distributional data corroborate the proposed DBSOM algorithms.
		\end{abstract}
		
	\end{frontmatter}
	
%	\linenumbers
	
	\section{Introduction}
	
Current big-data age requires the representation of voluminous data by summaries with loss of information as small as possible.
Usually, this is achieved by describing data subgroups according to descriptive statistics of their distribution (e.g.: the mean, the standard deviation, etc.)
Alternatively, when a dataset is observed with respect to a numerical variable, it can be described either by the estimate of the theoretical distribution that best fits the data or by an empirical distribution.

 In these cases, each set of observations is described by a distribution-valued data, and we call \textit{distributional-valued variable} a more general type of  variable whose values are one-dimensional empirical or estimated probability or frequency distributions on  numeric support.

\emph{Symbolic Data Analysis} \cite{BoDid00} introduced distributional-valued variables as particular set-valued variables, namely, modal variables having numeric support.
A particular type of distributional-valued variable is a histogram-valued variable, whose values are histograms.

%Thus, we call \textit{distributional-valued variable} a more general type of \sout{attribute} \textcolor{red}{variable} whose values are one-dimensional empirical or estimated probability or frequency distributions on a numeric support.

Such kind of data is arising in many practical situations.
 Official statistical institutes collect data about territorial units or administrations and often they carry out them as empirical distributions or histograms.
 Similarly, data are often available as aggregates in order to preserve individuals' privacy. For instance, bank transactions or measurements regarding patients of a hospital are often provided as histograms or empirical distributions rather than as individual records.

So far, several methods have been proposed for estimating distributions from a set of classical data, while few methods have been developed for the data analysis of objects described by distribution-valued variables. 

Among exploratory methods, Kohonen Self-Organizing Map (SOM) presents both visualization and clustering properties \cite{Kohonen01,Kohonen13}.
SOM is  based on a map of nodes (neurons) organized on a regular low dimensional grid where the neurons present a priori neighborhood relations.
Each neuron is described by a prototype vector (a model) and it is associated with a set of the input data.

In this sense, SOM carries out clustering while preserving the topological order of the prototype vectors on the map: the more the neurons are adjacent on the map, the more they are described by similar prototypes, whereas different prototypes are associated with neurons that are distant on the map.
Besides, during the training step, each object must be assigned to a neuron. This can be done in two ways. The classical SOM algorithm assigns an input data to the closest BMU (Best Matching Unit), namely the neuron that is described by the closest prototype. Following the approach considered by Kohonen \cite{Kohonen01} too, it is possible to consider the assignment as a part of an optimization problem. In this case, an objective function is associated with a SOM that is minimized according to the prototypes definition (a representation step) and the data assignment. Thus, the assignment is not merely done accordingly to the closest BMU but according to the BMU that allows a minimization of the objective function.
After that each object is assigned to the optimal BMU, the corresponding representative and a subset of representatives of its neighbors on the grid are modified aiming to fit better the data set.
%(namely, what we will call the criterion function in the following).

An important property of the SOM is that it preserves at the best the original topological structure of the data: the objects that are similar in the original space have their corresponding representatives similar and located close in the map.
Finally, the training of the SOM can be incremental or batch.
Kohonen \cite{Kohonen13} states that for practical applications, the batch version of the SOM is the more suitable. However, when the data is presented sequentially, as in stream data, the training has to be incremental.

SOM can be considered as a distance-based clustering method, thus the definition of a distance between the data is essential, mainly in our case where data are one dimensional distributions.

The literature on clustering distributions includes several proposals. In \cite{IrVer06} it is proposed a hierarchical clustering method which uses a Wasserstein distance for comparing distributions estimated by means of histograms. Using the same distance function  \cite{IRVERLEC06} propose a method based on the Dynamic Clustering Algorithm (DCA) \cite{DiSIM76}. The latter is a centroid based algorithm which generalizes the classic $k$-means.  It optimizes an internal homogeneity criterion by performing, iteratively, a representation and an allocation step until the convergence to a stable value of the optimized criterion. Another centroid based method has been introduced in \cite{Terada10}. It is a $k$-means algorithm which uses empirical joint distributions. Finally, \cite{VracEtAL12} propose a Dynamic Clustering Algorithm based on copula.
All the mentioned algorithms require an appropriate dissimilarity measure for comparing distributions. Among these, Wasserstein distances \cite{Rush01} have interesting features, as investigated in \cite{IrpVer2015}.

In classical clustering, usually, it is assumed that variables play the same role in the grouping process. Sometimes a standardization or a rescaling step is performed before running the clustering algorithm but this step does not assure that each variable participates to the clustering process according to its clustering ability. Indeed, there is a wide and unresolved debate about the variable transformation in clustering.
	The use of adaptive distances \citep{diday77} in clustering is a valid  approach for the identification of the importance of variables in the clustering process.
In the framework of Symbolic Data Analysis, several Dynamic Clustering algorithms including adaptive distances providing relevance weights for the variables, have been introduced for interval-valued data \cite{DeCDeS10,DECAYVES09,DECLEC09}, for histogram-valued data
\cite{Kim_Bill_2011,Kim_Bill_2013}  and for modal symbolic data \cite{Hardy2004,Korenjak-Cerne1998,KorenjakCerne2002}.
The use of adaptive distances, based on the $L_2$ Wasserstein metric, has been also proposed in the framework of Dynamic Clustering algorithm by \cite{IrpinoESWA}. Still, a fuzzy version of such algorithm is available in \cite{Irpino2017}.

Some extensions of the SOM have been proposed for interval-valued data \cite{Bock2002,Durso11,Cabanes13,Hajjar11a,Hajjar11b,Hajjar11c,Hajjar13}. Nonetheless, to the best of our knowledge, SOM algorithms for distributional data have not yet been proposed.
Moreover, usually SOM algorithms assume that the variables have the same importance for the training of the map, i.e, they assume that the variables have the same relevance.
However, in real applications some variables are irrelevant and some are less relevant than others \cite{diday77,friedman04,frigui04,huang05}.
%Furthermore, the relevance of each variable to each cluster may be different, i.e., each cluster may have a different set of relevant variables \cite{diday77,friedman04,frigui04,huang05}.
In order to consider the role of the variables in the partitioning structure, different approaches can be adopted.

A first strategy is to set the weight of variables according to the apriori knowledge about the application domain and, then, to perform the SOM procedure to train the map

%obtain a partition of the objects of the data set.
A suitable alternative is to add a step to the algorithm which computes, automatically, the weights for the variables. This approach has been used in Diday and Govaert \cite{diday77} which propose adaptive distances in which the automatically computed weights have a product to one constraint.

Recently, De Carvalho et al \cite{CarvalhoBS16} proposes batch SOM algorithms that learn a relevance weight for each interval-valued variable during the training phase of the SOM thanks to adaptive distances.
In particular, it is proposed to associate a relevance weight to each variable by introducing a weighting step in the algorithm and by modifying the optimized criterion function.

%    clustering process. Further, weights can be used for performing a variable selection in clustering process, see \cite{Modha2003}.
		
 %In the framework of k-means on classic scalar data with automatic relevance weights estimation a second approach was proposed in \cite{Huang_05}, where a sum to one constraint on the weights is considered. However, this algorithm needs to fix in advance (by the user or after  cross-validation) a smoothing parameter for the relevance weights.
	
	\subsection*{Main contributions}
The present paper extends the Batch Self Organized Map (BSOM) algorithm to histogram data. We refer to it as DBSOM, where D stands for Distributional data.
Since the DBSOM cost function depends on a distance between data that are distributions, among the different distances for comparing distributions, we propose the use of the Wasserstein distance. It belongs to a family of distance measures defined by different authors in a variety of contexts of analysis and
with different norms. %norm.
In our context, we consider the squared Wasserstein distance, also known as Mallows distance, as defined in \cite{Rush01}, and here named $L_2$ Wasserstein distance. The main motivation of this choice can be found in \cite{VerIRP08did}, where a comparison with other metrics is proposed. It is related to the possibility of defining barycenters of sets of distributions as Fr\'echt means, improving the ease of interpretation of the obtained results. 
Moreover, it allows of defining the variance over any distribution variables \cite{IrpVer2015}.

Besides, for the $L_2$ Wasserstein distance it has been proved an important decomposition in components related to the location (means) and variability and shape over the compared distributions \cite{IrpinoR07}.
Another contribution of the paper is the introduction of a system of weights for considering the different importance of the variables in the clustering process. Indeed, the classical BSOM assumes that each variable has the same role in the map learning, where standardization of the variables is usually performed before the algorithm starts (like in the Principal Component Analysis). However, the effect of teach variable can be relevant in the learning process. To take that into consideration, we introduce an \textit{Adaptive} version of DBSOM, denoted ADBSOM. We call it \textit{Adaptive} since it is based on the use of adaptive distances proposed by \cite{diday77} in clustering. As we will show, the optimization process of ADBSOM allows to compute automatically a system of weights for the variables.
Since the $L_2$ Wasserstein distance can be decomposed in two components related to the locations  (means), variability and shape, we also propose a system of weights for the two components of each distributional variable. This enriches the interpretation of the components of the variables that are relevant in the learning process and for the results.
The procedure is performed through an additional step of the ADBSOM algorithm that automatically learns relevance weights for each variable and/or for each component of the distance.
Some preliminary results were presented in \cite{IrpinoVC2012}.
In addition to extending the methods mentioned above, we propose new variants of the algorithm which consider new constraints in the optimization process.

The paper is organized as follows.
Sec. \ref{SEC_method} introduces details of our proposal focusing on the criterion function optimized by each algorithm; the computation of the relevance weights on data described by distributional-valued variables; the adaptive distances for distributional data. Sec. \ref{SEC_apply} provides an application of the proposed algorithms to real-world datasets. Sec. \ref{SEC_conl} concludes the paper with a discussion on the achieved results on the several datasets.
	
\section{Batch SOM algorithms for distributional-valued data} \label{SEC_method}

SOM is proposed as an efficient method
to address the problem of clustering and visualization, especially for high-dimensional data having as input a topological structure. Since a grid of neurons is
chosen for defining the topology of the map, the procedure of mapping a vector from
data space onto the map consists in finding the neurons with the closest weighted distance to the data space vector.
A correspondence between the input space and the mapping space is built such that, two close data in the input space, should activate the same neuron, or two neighboring neurons, of the SOM. A prototype describes each neuron.
Neighboring neurons in the map providing the Best Matching Unit (BMU) of a data update their prototypes to better represent the data.

An extension of SOM to histogram data is suitable to analyze data
that are already available in aggregate form, also generated as syntheses of a huge amount of original data.
This paper proposes batch SOM algorithms for distributional-valued data that automatically provides a set of relevance weights for the different variables.
We present two new batch SOM algorithms, namely DBSOM (Distributional Batch SOM) and ADBSOM (Adaptive Distributional Batch SOM). Both algorithms are based on the $L_2$ Wasserstein distance between distributional-valued data.
Specifically, during the training of the map, ADBSOM computes relevance weights for each distributional-valued variable.
Such relevance weights are assumed as parameters of the dissimilarity function that compares the prototypes and the data items.
Therefore, the computed values of such weights allow selecting the importance of the distributional-valued variables to the training of the map.
SOM provides both a visualization (given by the proximity between the neurons) and a partitioning of the input data (by an organization of the data in clusters).

\subsection{DBSOM criterion for distributional-valued data}

This section extends the classical objective function of batch SOM to the case of distributional-valued variables.

Let $E =\{e_1,\ldots,e_N\}$ be a set of $N$ objects described by $P$ distributional-valued variables $Y_j$ ($j=1, \ldots, P$).
Each $i$-th object $e_i  (1 \leq i \leq N)$ is represented by $P$ distributions (or distributional-valued data) $y_{ij} \, (1 \leq j \leq P)$. These are elements of an \textit{object vector} $\mathbf{y}_i = (y_{i1},\ldots,y_{iP})$. With each one-dimensional distributional data $y_{ij}$ is associated: a one-dimensional estimated density function $f_{ij}$, the corresponding cumulative distribution function (cdf) $F_{ij}$ and the quantile function (qf) $Q_{ij}=F_{ij}^{-1}$ (namely, the inverse of the cdf).

Therefore, the distributional-valued data set $\mathcal{D} = \{\mathbf{y}_1,\ldots,\mathbf{y}_N\}$ is collected in an \textit{object table}:

$$\mathbf{Y} = \big(y_{ij}\big)_{\substack{1 \leq i \leq N\\1 \leq j \leq P}}$$
\noindent where each element is a (one dimensional) distribution $y_{ij}$.

SOM is a low-dimensional (mainly, two-dimensional) regular grid, named {\em map}, that has $M$ nodes named {\em neurons}.
A SOM algorithm induces a partition where to each cluster corresponds a unique neuron described by prototype vector.
Thus, the neuron $m \, (1 \leq m \leq M)$ is associated with a cluster $C_m$ and a prototype \textit{object vector} $\mathbf{g}_m$.

The assignment function $f: \mathcal{D} \mapsto \{1,\ldots,M\}$ assigns an index $m = f(\textbf{y}_i) \in \{1,\ldots,M\}$ to each distributional-valued data  $\mathbf{y}_i$ according to:

\begin{equation} \label{c1}
 f(\mathbf{y}_i) = m = \argmin_{1 \leq r \leq M} d^T(\mathbf{y}_i, \mathbf{g}_r) \
\end{equation}

\noindent where $d^T$ is a suitable dissimilarity function between the object vectors $\mathbf{y}_i$ and the prototype vectors $\mathbf{g}_r$.

The partition $\mathcal{P} = \{C_1,\ldots,C_M\}$ carried out by SOM is obtained according to an assignment function that provides the index of the cluster of the partition to which the object $\mathbf{y}_i$ is assigned: $C_m = \{e_i \in E: f(\textbf{y}_i) = m\}$.

%The assignment function $f: \mathcal{D} \mapsto \{1,\ldots,M\}$ associates each data unit $\mathbf{y}_i$ to an index $m = f(\textbf{y}_m)$ of the index set $\{1,\ldots,M\}$, \textcolor{red}{where $\mathcal{D} = \{\mathbf{y}_1, \ldots, \mathbf{y}_N\}$ is a distributional-valued data set}.
%The partition $\mathcal{P} = \{C_1,\ldots,C_M\}$ associated with a SOM is defined by the assignment function which gives the index of the cluster of partition $\mathcal{P}$ to which the object $\mathbf{y}_i$ belongs to, $C_m = \{e_i \in E: f(\textbf{y}_i) = m\}$.

Since the variables are distributional-valued, each prototype $\mathbf{g}_m \, (1 \leq m \leq M)$ is a vector of $P$ distributional data, i.e., $\mathbf{g}_m = (g_{m1}, \ldots, g_{mP})$, where each $g_{mj} \, (1 \leq j \leq P)$ is a distribution.
Besides, with each one-dimensional distributional data $g_{mj}$ are associated: an estimate density function $f_{g_{mj}}$, the cdf $G_{mj}$ and the qf $Q_{g_{mj}}$\footnote{ We remark that $g_{mj}$ is a distributional data because the corresponding quantile function $Q_{g_{mj}}$ is a weighted average quantile function (for further details see \cite{Gilchrist} and \cite{IrpVer2015})}.
Hereafter, $\mathbf{G}$ is the matrix of the descriptions of each $g_{mj}$ associated with the prototypes:
$$\mathbf{G} =
%\big(\mathbf{g}_1, \ldots, \mathbf{g}_M \big)^{\prime} =
\big(g_{mj}\big)_{\substack{1 \leq m \leq M\\1 \leq j \leq P}}$$
\noindent where each cell contains a (one dimensional) distribution $g_{mj}$.

With the purpose that the obtained SOM represent the data set $\mathcal{D}$ accurately, the prototype matrix $\mathbf{G}$ and the partition $\mathcal{P}$ are computed iteratively according to the minimization of an suitable objective function (also known as \textit{energy function} of the map), hereafter refered as $J_{DBSOM}$, defined as the sum of the dissimilarities between the prototypes (best matching units) and the data unities:

\begin{equation}\label{crit-1}
J_{DBSOM}(\mathbf{G}, \mathcal{P}) = \sum_{i = 1}^N d^T(\mathbf{y}_i, \mathbf{g}_{f(\mathbf{y}_i)}) 	
\end{equation}

%%%%%%%%%%%%%%%%%%%%%%%%%%%%%%%%%%%%%%%%%%%%%%%%%%%%%%%%%%%%%%%%%%%%%%%%%%%%%%%%%%%%%%%%%%%%%%%%%%%%%%%%%%%%%%%%%%%%%%%%%%%%%%%%%%%%%%%%%%%%%%%%%%%%%%%%%%%%%%%%%%%%

Dissimilarities between each object and all the prototype vectors are needed to be computed. The best matching unit (BMU) is the winner neuron, i.e., the neuron indexed by $m = f(\mathbf{y}_i)$ with prototype vector $\mathbf{g}_{m}$ of minimum error.
%closest to the object.
The dissimilarity function, $d^T$, that is used to compare each object $\mathbf{y}_i$ with each prototype $\mathbf{g}_h$, is computed as follows:

\begin{equation}\label{cb-1}
d^T(\mathbf{y}_i, \mathbf{g}_m) = \sum_{h = 1}^M \mathcal{K}^T(d(m,h)) \, d^2_W(\mathbf{y}_i,\mathbf{g}_h)
\end{equation}

In equation (\ref{cb-1}), $d$ is the distance defined on the set of neurons. Usually, it is computed as the length of the shortest path on the grid between nodes (neurons) $m$ and $h$. $T$ is the {\em neighborhood radius}. $\mathcal{K}^T$ is the neighborhood kernel function that computes the neighborhood influence of winner neuron $m$ on neuron $h$. The neighborhood influence diminishes with $T$  \cite{Badran05}.

Since $\mathbf{y}_i$ is described by $P$ distributional variables as defined above, without any information about the multivariate distribution, we assume that $y_{i1}, \ldots y_{ij},$  $\ldots y_{iP}$ are marginal distributions. Thus, the (standard) L2 Wasserstein distance between the $i$-th object and the prototype $\mathbf{g}_h$ associated with the neuron $h$ is defined as follows:	
\begin{equation}\label{dimult}
		d^2_W\left(\mathbf{y}_i,\mathbf{g}_m\right)=\sum\limits_{j=1}^P d^2_W\left(y_{ij}, g_{mj}\right).
	\end{equation}

Following the \citeauthor{Rush01} \cite{Rush01} notation, the squared $L_2$ Wasserstein distance between $y_{ij}$ and $g_{mj}$ is defined as:	

	\begin{equation}\label{HOMSQ}
		d^2_W(y_{ij},g_{mj})=\int\limits_{0}^{1} {\left[ {Q _{ij} (p) - Q_{g_{mj}}(p)} \right] ^2 dp}.
	\end{equation}

In \cite{Irpino2017} is shown that the squared $L_2$ Wasserstein distance presents more interpretative properties compered with other distances between distributions. Especially it can be decomposed in two independent distance-components as follows:
	\begin{equation} \label{eq:IrpRoma2}
 d^{2}_{W}(y_{ij},g_{mj})=(\bar{y}_{ij}-{\bar y}_{g_{mj}})^{2}+\int\limits_{0}^{1} {\left[ {Q^c _{ij} (p) - Q^c_{g_{mj}}(p)} \right] ^2 dp},
	\end{equation}
where: $\bar{y}_{ij}$ and ${\bar y}_{g_{mj}}$ are the means;  $Q^c_{ij}(p)=Q_{ij}(p)-\bar{y}_{ij}$ and $Q^c_{g_{mj}}(p)=Q_{g_{mj}}(p)-\bar{y}_{g_{mj}}$ are the centered quantile functions of $y_{ij}$ and $g_{mj}$, respectively. Briefly, the squared $L_2$ Wasserstein distance is expressed as the sum of the squared Euclidean distance between means and the squared $L_2$ Wasserstein distance between the centered quantile functions corresponding to the two distributions. We rewrite the same equation as follows:

\begin{equation} \label{eq:IrpRoma3}
 d^{2}_{W}(y_{ij},g_{mj})=d^{2}(\bar{y}_{ij},{\bar y}_{g_{mj}})+d^{2}_{W}(y_{ij}^c,g_{mj}^c).
\end{equation}

Finally, the standard multivariate squared $L_2$ Wasserstein distance between the $i$-th object $\mathbf{y}_i$ and the prototype $\mathbf{g}_m$ is as follows
		\begin{equation}
			\label{dist-1}
			d^2_W(\mathbf{y}_i,\mathbf{g}_m) = \sum\limits^{P}_{j=1}({\bar y}_{ij}-{\bar y}_{g_{mj}})^{2}+ \sum\limits^{P}_{j=1}d^2_W(y^c_{ij},g^c_{mj}).
		\end{equation}
		
		For the rest of the paper, we denote with $dM_{im,j}=({\bar y}_{ij}-{\bar y}_{g_{mj}})^{2}$
        %$dM_{hk,j}=({\bar y}_{kj}-{\bar g}_{{hj}})^{2}$
        the squared Euclidean distance between the means of distributional data $y_{ij}$ and ${g_{mj}}$, and with $dV_{im,j}=d^2_W(y^c_{ij},g^c_{mj})$ the squared $L_2$ Wasserstein distance between the centered distributional data. Equation (\ref{dist-1}) can be written in a compact form as follows:
		\begin{equation}
			\label{dist-1C}
			d^2_W(\mathbf{y}_i,\mathbf{g}_m) = \sum\limits^{P}_{j=1} (dM_{im,j}+dV_{im,j}).
		\end{equation}

Therefore, the generalized distance $d^T(\mathbf{y}_i, \mathbf{g}_{f(\mathbf{y}_m)})$ (equation \ref{cb-1}) is a weighted sum of the non-adaptive multivariate squared $L_2$ Wasserstein distances $d^2_W$ computed between the vector $\mathbf{y}_i$ and the prototypes of the neighborhood of the winner neuron $f(\textbf{y}_m)$.

\subsection{Adaptive DBSOM criterion for distributional-valued data}

Usually, SOM models assume that the variables have the same importance for the clustering and visualization tasks. Nonetheless, in real applications, some variables are irrelevant and others are more or less relevant.
Moreover, each cluster may have its specific set of relevant variables \cite{diday77,friedman04,frigui04,huang05}.

In the framework of clustering analysis, adaptive distances \cite{diday77} have been proposed for solving the issue. Adaptive distances are weighted distances, where, generally, a positive weight is associated with each variable according to its relevance in the clustering process, such that the system of weights satisfies suitable constraints.

Adaptive distances were originally proposed in a k-means-like algorithm in two different ways. In a first algorithm, a weight is associated with each variable for the whole dataset (we call it a Global approach). In a second approach, considering a partition of the dataset into $M$ cluster, a weight is associated with each variable and each cluster (we call it a Cluster-wise approach). In this paper, we remark that each neuron of the SOM is related to a Voronoi set that can be considered as a cluster of input data. Similarly to \cite{IrpinoVC2012}, in this paper, we extend this idea to a dataset of distributional data. Following the same approach, we go beyond by exploiting the decomposition of the squared L2 Wasserstein distance in Eq. (\ref{dist-1C}), proposing to weight the components too. Namely, we provide a method for observing the relevance of the two aspects of a distributional variable related to the two components. The current proposal differs from the one in \cite{IrpinoVC2012} for the constraints on the weights.

In this paper, we denote with $\boldsymbol{\Lambda}$ the matrix of relevance weights.
The dimension of $\boldsymbol{\Lambda}$ is $P\times 1$ or, respectively,  $2P\times 1$, if relevance weights are associated with each variable or, respectively, each component for the whole dataset;
%and
$P\times M$ or, respectively,  $2P\times M$ for each variable or, respectively, each component for each neuron.
We recall that adaptive distances rely on relevance weights that are not defined in advance, but they depend on the minimization of the objective function, here denoted with $J_{ADBSOM}$, which measures the dispersion of data around the prototypes. Obviously, the trivial solution of such minimization is obtained when $\mathbf\Lambda$ is a null matrix. 

To avoid the trivial solution, a constraint on the relevance weights is necessary. In the literature, a constraint on the product \cite{diday77} or on the sum  \cite{Huang_05}, usually to one, is suggested.		 
Even if the latter approach appears more natural, it relies on the tuning of a parameter that must be fixed in advance. So far, a consensus on an optimal value is still missing, we do not discuss its use in this paper.

In the framework of clustering analysis for non-standard data, adaptive squared  $L_2$ Wasserstein distances were applied (see \cite{DeCDeS10,DECLEC09}) for deriving relevance weights for each variable and cluster.
%In those approaches, the weights satisfy a sum or a product-to-one constraints.
\citeauthor{IrpinoESWA} \cite{IrpinoESWA} provided a component-wise adaptive distance approach to clustering, but the relevance weights are related to two independent constraints for each component of the distance, forcing the assignment of high relevance weights to components whose contributes to the clustering process are low too. 

In this paper, we propose the approach suggested in \cite{Irpino2017} that solve this issue.
Differently to the method proposed by \citeauthor{Kohonen01}\cite{Kohonen01}, we consider the training of the SOM as a set of iterative steps that minimizes a criterion function $J_{ADBSOM}$ \cite{Badran05}. The main difference is the allocation of objects to the Voronoi set of each neuron. Indeed, in the original formulation, that is the widely used one, the allocation is performed according to the minimum distance between the object and the prototype only. That approach does not guarantee a monotonic decreasing of the criterion function along the training step. In our case, at each step of the training of the SOM, a set $\mathbf{G}$ of $M$ prototypes, namely, a prototype for each neuron, the matrix $\mathbf\Lambda$ of relevance weights and the partition $\mathcal{P}$ of the input objects are derived by the the minimization of the error function $J_{ADBSOM}$ (know also as energy function of the map), computed as the following dispersion criterion:
\begin{equation}\label{crit-1-1}
J_{ADBSOM}(\mathbf{G}, \mathbf{\Lambda}, \mathcal{P}) = \sum_{i = 1}^N d^T_{\mathbf{\Lambda}}(\mathbf{y}_i, \mathbf{g}_{f(\mathbf{y}_i)}) 	
\end{equation}

The dissimilarity function, $d^T_{\mathbf{\Lambda}}$, that compares each data unit $\mathbf{y}_i$ to each prototype $\mathbf{g}_h \, (1 \leq h \leq M)$ is defined as:

\begin{equation}\label{cb-2}
d^T_{\mathbf{\Lambda}}(\textbf{y}_i, \textbf{g}_m) = \sum_{h = 1}^M \mathcal{K}^T(d(m,h)) \, d_{\mathbf{\Lambda}}(\mathbf{y}_i,\mathbf{g}_h)
\end{equation}

\noindent where $d$, $T$ and $\mathcal{K}^T$ are defined as in equation (\ref{cb-1}), while $d_{\mathbf{\Lambda}}(\mathbf{y}_i,\mathbf{g}_h)$ is one of the four following equations depending on the global or cluster-wise, variable or component assignment of the relevance weights:

			\begin{equation}\label{dist-1-1}
				d_{\mathbf{\Lambda}}(\mathbf{y}_i,\mathbf{g}_m)				= \sum_{j=1}^P  \lambda_j \,\left( dM_{im,j}+dV_{im,j}\right)
			\end{equation}
			\begin{equation}\label{dist-1-3}
				d_{\mathbf{\Lambda}}(\mathbf{y}_i,\mathbf{g}_m)
				= \sum_{j=1}^P \left(\lambda_{j,\mathcal{M}} \,dM_{im,j} + \lambda_{j,\mathcal{V}} \,dV_{im,j}\right)
			\end{equation}
			\begin{equation}\label{dist-1-2}
				d_{\mathbf{\Lambda}}(\mathbf{y}_i,\mathbf{g}_m)
				= \sum_{j=1}^P \lambda_{mj} \, \left(dM_{im,j} + dV_{im,j}\right)
			\end{equation}
			\begin{equation}\label{dist-1-4}
				d_{\mathbf{\Lambda}}(\mathbf{y}_i,\mathbf{g}_m)
				= \sum_{j=1}^P \left(\lambda_{mj,\mathcal{M}} \, dM_{im,j} + \lambda_{mj,\mathcal{V}} \,dV_{im,j}\right)
			\end{equation}

Thus, the generalized distance $d^T_{\mathbf{\Lambda}}(\textbf{y}_i, \textbf{g}_{f(\textbf{y}_i)})$ is a weighted sum of the adaptive multivariate squared $L_2$ Wasserstein distances $d_{\mathbf{\Lambda}}(\mathbf{y}_i,\mathbf{g}_m)$ computed between the vector $\mathbf{y}_i$ and the prototype of the neighborhood of the winner neuron $f(\textbf{y}_m)$.

For avoiding the trivial solution (namely, null $\lambda$'s), we use the above mentioned product constraint.
We suggest four different constraints on the relevance weights, that is:
%either
%, or to be greater or equal to zero and to satisfy the sum-to-one constraint.
		\begin{description}
			\item[(P1) A product constraint  for Eq. (\ref{crit-1-1}) and Eq. (\ref{dist-1-1})] See \cite{diday77}:
			\begin{equation}\label{Con-P-1}
				\prod_{j=1}^P \lambda_{j} = 1, \; \; \;
                \lambda_{j} > 0
			\end{equation}

			\item[(P2) A product constraints for Eq. (\ref{crit-1-1}) and Eq. (\ref{dist-1-3})] See \cite{IrpinoESWA} and \cite{DeCA_IR_VE_IEEE_2015}:
			\begin{equation}\label{Con-P-3}
				\prod_{j=1}^P \left(\lambda_{j,\mathcal{M}} \cdot \lambda_{j,\mathcal{V}}\right)= 1, \, \; \; \; \lambda_{j,\mathcal{M}} > 0,\, \; \; \; \lambda_{j,\mathcal{V}} > 0
			\end{equation}

			\item[(P3) $M$ product constraints for Eq. (\ref{crit-1-1}) and Eq. (\ref{dist-1-2})] See \cite{diday77}:
			\begin{equation}\label{Con-P-2}
				\prod_{j=1}^P \lambda_{mj} = 1, \, \; \; \;  \lambda_{mj} > 0\quad m=1,\ldots,M;
			\end{equation}

			\item[(P4) $M$ product constraints for Eq. (\ref{crit-1-1}) and Eq. (\ref{dist-1-4})]See \cite{DeCA_IR_VE_IEEE_2015} and \cite{IrpinoESWA}:
			\begin{equation}\label{Con-P-4}
				\prod_{j=1}^P \left( \lambda_{mj,\mathcal{M}}\cdot \lambda_{mj,\mathcal{V}}\right) = 1, \, \; \; \; \lambda_{mj,\mathcal{M}} > 0, \; \; \;
                \, \lambda_{mj,\mathcal{V}} > 0\quad m=1,\ldots,M;
			\end{equation}
\end{description}

\noindent \emph{Remark.} Note that the weights of each variable, or of each component in the cluster-wise scheme, are locally estimated. 
This means that at each iteration the weights change. Moreover, each neuron (whose Voronoi set represent a cluster) has its specific set of weights.
On the other hand, weights defined according a global scheme are the same for all the clusters.
In general, note that a relevant variable (or component) has a weight greater than 1 because of the product-to-one constraint.

\subsection{The batch SOM optimization algorithm for distributional-valued data}

In this paper, we use a batch training of the SOM, namely, all the data are presented to the map at the same time. Once $T$ (namely, the neuron radius) is fixed, the training of the DBSOM depends on the minimization the criterion function $J_{DBSOM}$ which is based on classical squared $L_2$ Wasserstein distances. Thus no relevance weights are computed. The training task alternates a representation and an assignment step iteratively.
The representation step returns the optimal solution for the prototypes describing the neuron of the map. In the assignment step objects are optimally allocated to the Voronoi sets of each neuron of the map.
Differently, the ADBSOM algorithm is trained through the minimization of $J_{ADBSOM}$ function. In that case, three steps are iterated:  a representation, a weighting and an assignment one. The representation and assignment steps are performed like in DBSOM. The new weighting step carries out optimal solutions for  relevance weights according one of the four proposed schemes.

\subsubsection{Representation step}\label{s-rep}

This section focuses on the optimal solution, for a fixed radius $T$, of the prototype of the cluster associated to each neuron during the training of the DBSOM and ADBSOM algorithms.

In the representation step of DBSOM, for a fixed partition $\mathcal{P}$, the objective function $J_{DBSOM}$ is minimized regarding to the components of the matrix $\mathbf{G}$ of prototypes.
%according to the minimum distance to the the prototypes (in the matrix $\mathbf{G}$).

Similarly, in the representation step of ADBSOM, for a fixed partition $\mathcal{P}$ and for a fixed set of weights in the matrix  $\mathbf{\Lambda}$, the objective function $J_{ADBSOM}$ is minimized regarding to the components of the matrix $\mathbf{G}$ of prototypes.
%according to the minimum distance to the the prototypes (in the matrix $\mathbf{G}$).

DBSOM looks for the prototype $\textbf{g}_m$ of the cluster $C_m \, (1,\leq m \leq M)$  that minimizes the following expression:
\begin{equation}\label{crit_1}
\sum_{j=1}^{P} \sum_{i=1}^{N} \mathcal{K}^T(d(f(\textbf{y}_i), m)) \; d^2_W(\mathbf{y}_i,\mathbf{g}_m)
\end{equation}

ADBSOM looks for the prototype $\textbf{g}_m$ of the cluster $C_m \, (1,\leq m \leq M)$  that minimizes the following expression:
\begin{equation}\label{crit_2}
\sum_{j=1}^{P} \sum_{i=1}^{N} \mathcal{K}^T(d(f(\textbf{y}_i), m)) \; d_{\mathbf{\Lambda}}(\mathbf{y}_i,\mathbf{g}_m)
\end{equation}
Since both problems are additive, and depends on quadratic terms, the  description of the prototypes $g_{mj}$ as distributions for each variable ($m=1,\ldots,M$, $j=1,\ldots,P$), is obtained as solution of the following minimization problem:

\begin{equation}\label{eq-prot-opt}
\sum_{i=1}^{N}  \mathcal{K}^T(d(f(\textbf{y}_i), m)) \; \left[({\bar y}_{ij}-{\bar y}_{g_{mj}})^{2} + d^2_W(y^c_{ij},g^c_{mj})\right] \longrightarrow \mbox{ Min .}
\end{equation}
Thus, for each cluster,  setting to zero the partial derivatives w.r.t. ${\bar y}_{g_{mj}}$ and $Q^c_{g_{mj}}$ \cite{IrpVer2015}, the quantile function of the probability density function (\emph{pdf}) describing $g_{mj}$ is obtained as follows: 

\begin{equation}\label{prot-1}
			Q_{g_{mj}}=Q^{c}_{g_{mj}}+{\bar y}_{{g_{mj}}}=
			\frac{
				\sum\limits_{i=1}^N \mathcal{K}^T(d(f(\mathbf{y}_i),m) \, Q^{c}_{ij}
			}
			{
				\sum\limits_{i=1}^N \mathcal{K}^T(d(f(\mathbf{y}_i),m)
			}
			+
			\frac{
				\sum\limits_{i=1}^N \mathcal{K}^T(d(f(\mathbf{y}_i),m) {\bar y}_{ij}
			}
			{
				\sum\limits_{i=N}^n \mathcal{K}^T(d(f(\mathbf{y}_i),m)
			}.
\end{equation}

\subsubsection{Weighting step}\label{s-weight}

The aim of this section is to provide, for a fixed radius $T$, the optimal solution  of the relevance weights of the distributional-valued variables during the training of the ADBSOM algorithms.

In the weighting step of ADBSOM, fixed the partition $\mathcal{P}$ and the prototypes in the matrix  $\mathbf{G}$, the objective function $J_{ADBSOM}$ is minimized w.r.t. to the weights, elements of the matrix of $\mathbf{\Lambda}$.

		\begin{proposition}\label{prop-weight-1}
		    The relevance weights are calculated depending upon the adaptive squared $L_2$ Wasserstein distance:
			\begin{description}
				
				%\begin{enumerate}
				\item[(P1)]$$\mathit{If}\;J_{ADBSOM}(\mathbf{G}, \mathbf{\Lambda}, \mathcal{P}) = \sum_{m=1}^M \sum_{i=1}^N \sum_{j=1}^P  \mathcal{K}^T(d(f(\mathbf{y}_i),m)) \,\lambda_j \, d^2_W(y_{ij},g_{mj})$$  subject to $\prod_{j=1}^P \lambda_{j} = 1, \, \lambda_{j} > 0$, then $P$ relevance weights are derived as follows:
				\begin{equation}
					\label{W-Glo-1V-Prod}
					\lambda_j=\frac{{{{\left\{ {\prod\limits_{r = 1}^P \left[{\sum\limits_{m = 1}^M {\sum\limits_{i = 1}^N {{\mathcal{K}^T(d(f(\mathbf{y}_i),m))}{d^2_W\left( {{y_{ir}},{g_{mr}}} \right)}} } } \right]} \right\}}^{\frac{1}{P}}}}}{{\sum\limits_{m = 1}^M\sum\limits_{i = 1}^N {{\mathcal{K}^T(d(f(\mathbf{y}_i),m))}{d^2_W \left({y_{ij},g_{mj}} \right)}} }}
				\end{equation}

				\item[(P2)] $$\mathit{If}\; J_{ADBSOM}(\mathbf{G}, \mathbf{\Lambda}, \mathcal{P}) = \sum_{m=1}^M \sum_{i=1}^N\sum_{j=1}^P  \mathcal{K}^T\left(d(f(\mathbf{y}_i),m)\right) (\lambda_{j,\mathcal{M}}dM_{im,j} + \lambda_{j,\mathcal{V}}dV_{im,j})$$
				subject to $\prod_{j=1}^P \left(\lambda_{j,\mathcal{M}}\cdot \lambda_{j,\mathcal{V}}\right) = 1$, $\lambda_{j,\mathcal{M}} > 0$ and  $\lambda_{j,\mathcal{V}} > 0$, then $2 \times P$ relevance weights are derived as follows:
				
				\begin{eqnarray}
					\label{W-Glo-2C-Prod}
					\lambda_{j,\mathcal{M}}=\frac{
                    {{{{
                    \left\{
                    {
                    \prod\limits_{r = 1}^P {
                    \left[
                    \sum\limits_{m = 1}^M {
                    \sum\limits_{i = 1}^N {{\mathcal{K}^T(d(f(\mathbf{y}_i),m))}d{M_{im,r}}}
                      }
                     \right]
                    }
                    }
                    {
                    \left[
                    \sum\limits_{m = 1}^M {
                    \sum\limits_{i = 1}^N {{\mathcal{K}^T(d(f(\mathbf{y}_i),m))}d{V_{im,r}}}
                      }
                     \right]
                    }
                    \right\}
                    ^{\frac{1}{2P}}}}}}
                    }{{\sum\limits_{m = 1}^M\sum\limits_{i = 1}^N {{\mathcal{K}^T(d(f(\mathbf{y}_i),m))}d{M_{im,j}}} }},\;\text{and}\;\nonumber\\
					\lambda_{j,\mathcal{V}}=\frac{
                    {{{{
                    \left\{
                    {
                    \prod\limits_{r = 1}^P {
                    \left[
                    \sum\limits_{m = 1}^M {
                    \sum\limits_{i = 1}^N {{\mathcal{K}^T(d(f(\mathbf{y}_i),m))}d{M_{im,r}}}
                      }
                     \right]
                    }
                    }
                    {
                    \left[
                    \sum\limits_{m = 1}^M {
                    \sum\limits_{i = 1}^N {{\mathcal{K}^T(d(f(\mathbf{y}_i),m))}d{V_{im,r}}}
                      }
                     \right]
                    }
                    \right\}
                    ^{\frac{1}{2P}}}}}}
                    }{{\sum\limits_{m = 1}^M\sum\limits_{i = 1}^N {{\mathcal{K}^T(d(f(\mathbf{y}_i),h))}d{V_{im,j}}} }}.
				\end{eqnarray}

				\item[(P3)] $$\mathit{If} \;J_{ADBSOM}(\mathbf{G}, \mathbf{\Lambda}, \mathcal{P}) = \sum_{m=1}^M \sum_{i=1}^N\sum_{j=1}^P  \mathcal{K}^T(d(f(\mathbf{y}_i),m)) \,\lambda_{mj}d^2_W(y_{ij},g_{mj})
				$$ subject to $M$ constraints $\prod_{j=1}^P \lambda_{mj} = 1, \, \lambda_{mj} > 0$, then $M\times P$ relevance weights are derived as follows:
				\begin{equation}
					\label{W-Loc-1V-Prod}
					\lambda_{mj}=\frac{{{{ \left\{{\prod\limits_{r = 1}^P \left[{\sum\limits_{i = 1}^N {{\mathcal{K}^T(d(f(\mathbf{y}_i),m))}{d^2_W}\left( {{y_{ir}},{g_{mr}}} \right)} } \right]} \right\} }^{\frac{1}{P}}}}}{{\sum\limits_{i = 1}^N {{\mathcal{K}^T(d(f(\mathbf{y}_i),m))}{d^2_W}\left( {{y_{ij}},{g_{mj}}} \right)} }}
				\end{equation}

				\item[(P4)]$$\mathit{If}\;
				J_{ADBSOM}(\mathbf{G}, \mathbf{\Lambda}, \mathcal{P}) = \sum_{m=1}^M \sum_{i=1}^N\sum_{j=1}^P  \mathcal{K}^T(d(f(\mathbf{y}_i),m)) \,[\lambda_{ij,\mathcal{M}}dM_{im,j}+\lambda_{hj,\mathcal{V}} dV_{im,j})]$$ subject to $M$ constraints $\prod_{j=1}^P \left(\lambda_{mj,\mathcal{M}}\cdot\lambda_{mj,\mathcal{V}}\right) = 1$, $\lambda_{mj,\mathcal{M}} > 0$ and $\lambda_{mj,\mathcal{V}} > 0$,
				then $2 \times M \times P$ relevance weights  are derived as follows:
				
				\begin{eqnarray}
					\label{W-Loc-2C-Prod}
					\lambda_{mj,\mathcal{M}}=
                    \frac{
                    {
                    {
                    \left\{
                    {\prod\limits_{r = 1}^P \left[{\sum\limits_{i = 1}^N {{\mathcal{K}^T(d(f(\mathbf{y}_i),m))} \, dM_{im,r}} } \right]
                    \left[{\sum\limits_{i = 1}^N {{\mathcal{K}^T(d(f(\mathbf{y}_i),m))} \, dV_{im,r}} } \right]
                    }
                    \right\}^{\frac{1}{2P}}}}
                    }
                    {{\sum\limits_{i = 1}^N {{\mathcal{K}^T(d(f(\mathbf{y}_i),m))} \, dM_{im,j}} }}\quad \text{and} \,\nonumber\\
					\lambda_{mj,\mathcal{V}}=
                    \frac{
                    {
                    {
                    \left\{
                    {\prod\limits_{r = 1}^P \left[{\sum\limits_{i = 1}^N {{\mathcal{K}^T(d(f(\mathbf{y}_i),m))} \, dM_{im,r}} } \right]
                    \left[{\sum\limits_{i = 1}^N {{\mathcal{K}^T(d(f(\mathbf{y}_i),m))} \, dV_{im,r}} } \right]
                    }
                    \right\}^{\frac{1}{2P}}}}
                    }
                    {{\sum\limits_{i = 1}^N {{\mathcal{K}^T(d(f(\mathbf{y}_i),m))} \, dV_{im,j}} }}.\quad
				\end{eqnarray}
							
			\end{description}
		\end{proposition}

		\begin{proof}
		In the weighing step, we assume that the prototypes $\mathbf{G}$ and the partition $\mathcal{P}$ is fixed. The matrix of relevance weights $\boldsymbol{\Lambda}$ are obtained according to one of the four above mentioned constraints.
			The minimization of $J_{ADBSOM}$ is done by the Lagrange multipliers method. The four constraints allow for the following Lagrangian equations:
			\begin{align}
				\mathbf{(P1):}\;{\mathcal L} =& J_{ABSOM}(\mathbf{G}, \mathbf{\Lambda}, \mathcal{P})-\theta \left(\prod_{j=1}^P \lambda_{j} - 1 \right);\\
				\mathbf{(P2):}\;{\mathcal L} =& J_{ABSOM}(\mathbf{G}, \mathbf{\Lambda}, \mathcal{P})-\theta \left[\prod_{j=1}^P \left(\lambda_{j,\mathcal{M}}\cdot \lambda_{j,\mathcal{V}}\right) - 1 \right];\\
				\mathbf{(P3):}\;{\mathcal L} = &J_{ABSOM}(\mathbf{G}, \mathbf{\Lambda}, \mathcal{P})-\sum_{m=1}^M \theta_m \left(\prod_{j=1}^P \lambda_{ij} - 1\right);\\
				\mathbf{(P4):}\;{\mathcal L} =& J_{ABSOM}(\mathbf{G}, \mathbf{\Lambda}, \mathcal{P}) -\sum_{m=1}^M \theta_{m} \left[\prod_{j=1}^P \left(\lambda_{mj,\mathcal{M}}\cdot \lambda_{mj,\mathcal{V}}\right) - 1\right].
			\end{align}

			Setting to zero the partial derivatives of $\mathcal{L}$ with respect to the {$\lambda$'s} and the {$\theta$'s} respectively, a system of equations of the first order condition is obtained and their solution corresponds to the elements of the matrix $\mathbf{\Lambda}$.
            \end{proof}	

\subsubsection{Assignment step}\label{s-affect}

The aim of this section is to give the optimal partition of the clusters associated to the neurons of the SOM  in the assignment step of the DBSOM and ADBSOM algorithms.

Fixed the prototypes, elements of the  matrix $\mathbf{G}$, the objective function $J_{DBSOM}$ is minimized w.r.t. the partition $\mathcal{P}$ and each data unit \textbf{y}$_i$ is assigned to its nearest prototype (BMU).

\begin{proposition}\label{prop-part-1}
The objective function $J_{DBSOM}$ is minimized w.r.t. the partition $\mathcal{P}$ when the clusters $C_m \, (m=1,\ldots,M)$ are computed as:
\begin{equation} \label{af-l1}
C_m = \{\mathbf{y}_i \in {\mathcal{D}}: f(\mathbf{y}_i) = m = \argmin_{1 \leq r \leq M} d^T(\mathbf{y}_i, \mathbf{g}_r) \}
\end{equation}
\end{proposition}

\begin{proof}
Because the matrix of prototypes $\mathbf{G}$ is fixed, the objective function $J_{DBSOM}$ can be rewrite as:
$$
J_{DBSOM}(\mathcal{P}) = \sum_{i = 1}^N d^T(\mathbf{y}_i, \mathbf{g}_{f(\mathbf{y}_i)})
$$
Remark that if, for each $\mathbf{y}_i \in {\mathcal{D}}$, $d^T(\mathbf{y}_i, \mathbf{g}_{f(\mathbf{y}_i)})$ is minimized, then the criterion $J_{DBSOM}(\mathcal{P})$ is also minimized.
The matrix of prototypes $\mathbf{G}$ being fixed,
$d^T(\mathbf{y}_i, \mathbf{g}_{f(\mathbf{y}_i)})$ is minimized if $f(\mathbf{y}_i) = m = \argmin_{1 \leq r \leq M} d^T(\mathbf{y}_i, \mathbf{g}_r)$, i.e., if  $C_m = \{\mathbf{y}_i \in \mathcal{D}: f(\mathbf{y}_i) = m = \argmin_{1 \leq r \leq M} d^T(\mathbf{y}_i, \mathbf{g}_r) \} \, (m=1,\ldots,M)$.
\end{proof}

In the assignment step of ADBSOM, for a fixed matrix of prototype $\mathbf{G}$ and matrix of weights $\mathbf{\Lambda}$, the objective function $J_{ADBSOM}$ is minimized w.r.t. the partition $\mathcal{P}$ and each data unit  $\mathbf{y}_i$ is assigned to its nearest prototype.

\begin{proposition}\label{prop-part-2}
The objective function $J_{ADBSOM}$ is minimized w.r.t. the partition $\mathcal{P}$ when the clusters $C_m \, (m=1,\ldots,M)$ are computed as:
\begin{equation} \label{af-l2}
C_m = \{\textbf{y}_i \in \mathcal{D}: m = f (\mathbf{y}_i)= \argmin_{1 \leq r \leq M} d^T_{\mathbf{\Lambda}}(\textbf{y}_i, \textbf{g}_r) \}
\end{equation}
\end{proposition}

\begin{proof}
The proof is similar to the proof of the Proposition (\ref{prop-part-1}).
\end{proof}

\subsubsection{DBSOM and ADBSOM algorithms} \label{s-algo}

Algorithm \ref{alg1} summarizes the batch SOM algorithms DBSOM and ADBSOM for distributional-valued data.
In the initialization step, first the matrix of prototypes $\mathbf{G}$ is initialized by randomly choosing $m$ objects of the distributional-valued data-set and then, the weights of the matrix of relevance weights $\mathbf{\Lambda}$ are set to 1 (each component or each variable are assumed to have the same relevance). Besides, using the initialization of the matrix of prototypes $\mathbf{G}$ and the initialization of the matrix of relevance weights $\mathbf{\Lambda}$, the rest of the objects are assigned to the cluster represented by the nearest representative (BMU) to produce the initial partition $\mathcal{P}$. Finally, from the initialization of $\mathbf{G}$, $\mathbf{\Lambda}$ and $\mathcal{P}$, and with $T \leftarrow T_{max}$, the algorithm calls the ITERATIVE-FUNCTION-1 that provides the initial SOM map and the corresponding new initial values of $\mathbf{G}$, $\mathbf{\Lambda}$ and $\mathcal{P}$.

In the iterative steps, in each iteration a new radius $T$ is computed and for this new radius the algorithm call the ITERATIVE-FUNCTION-1 that alternates once two (DBSOM) or three steps (ADBSOM) aiming to provide the update of the matrix of prototypes $\mathbf{G}$, the matrix of relevance weights $\mathbf{\Lambda}$ and the partition $\mathcal{P}$.

In the final iteration, when the radius $T$ is equal to $T_{min}$,
only few neurons belong to the neighborhood of the winner neurons and the SOM algorithm behaves similar to k-means. 
The algorithm calls the ITERATIVE-FUNCTION-2 that alternates once two (DBSOM) or three steps (ADBSOM) until the assignments no longer change. The final iteration provides the final update of the matrix of prototypes $\mathbf{G}$, the matrix of relevance weights $\mathbf{\Lambda}$ and the partition $\mathcal{P}$.

\begin{algorithm}
\caption{DBSOM and ADBSOM algorithms} % give the algorithm a caption
\label{alg1} % and a label for \ref{} commands later in the document
\begin{algorithmic}[1]
\Require
\State the distributional-valued data set $\mathcal{D} = \{\mathbf{y}_1,\ldots,\mathbf{y}_N\}$;
the number $M$ of neurons (clusters); the size map; the kernel function $\mathcal{K}^T$; the dissimilarity $d$; the initial radius $T_{max}$ and the final radius $T_{min}$; the number $N_{iter}$ of iterations.
\Ensure
\State the SOM map; the partition $\mathcal{P}$; the matrix $\mathbf{G}$ of prototypes; the matrix $\mathbf{\Lambda}$ of weights.   
\State \textbf{INITIALIZATION:}
\State Set $t\leftarrow 0$ and $T \leftarrow T_{max}$;
\State Initialization the matrix of prototypes $\mathbf{G}^{(0)}$: select randomly $M$ distinct prototypes $\mathbf{g}_r^{(0)} \in \mathcal{D} \; (r=1,\ldots,M)$;
%\sout{Randomly select $M$ distinct prototypes $\mathbf{g}_r^{(0)} \in \mathcal{D} \; (r=1,\ldots,M)$ to initialize the matrix of prototypes $\mathbf{G}^{(0)}$;}
\State Initialization of the matrix of relevance weights: set $\mathbf{\Lambda}^{(0)}=\mathbf{1}$;
%\sout{Set $\mathbf{\Lambda}^{(0)}=\mathbf{1}$ to initialize the matrix of relevance weights;}
\State Initialization of the partition $\mathcal{P}^{(0)}$: assign each object $\mathbf{y}_i$ to the closest prototype $\mathbf{g}_r$ (BMU) according to $r = f^{(0)} (\mathbf{y}_i)= \argmin_{1 \leq m \leq M} d^T(\mathbf{y}_i, \mathbf{g}^{(0)}_m)$;
%\sout{Assign each object $\mathbf{y}_i$ to the closest prototype $\mathbf{g}_r$ (BMU) according to $r = f^{(0)} (\mathbf{y}_i)= \argmin_{1 \leq m \leq M} d^T(\mathbf{y}_i, \mathbf{g}^{(0)}_m)$, to initialize the initial partition $\mathcal{P}^{(0)}$;}
\State \Call{ITERATIVE-FUNCTION-1}{$\mathcal{D}$, $t=0$, $\mathbf{G}^{(0)}$, $\mathbf{\Lambda}^{(0)}$, $\mathcal{P}^{(0)}$, $T$}
\State \textbf{ITERATIVE STEPS:}
\Repeat
\State Set $t \leftarrow t + 1$; Compute $T=T_{Max} \left(\frac{T_{min}}{T_{Max}}\right)^{\frac{t}{N_{iter}}}$;
%$T=T_{max} (\frac{T_{min}}{T_{max}})^{\frac{t}{N_{iter}-1}}$;
%\vspace*{-0.15cm}
\State Set  $\mathcal{P}^{(t)} \leftarrow \mathcal{P}^{(t-1)}$, $\mathbf{\Lambda}^{(t)} \leftarrow \mathbf{\Lambda}^{(t-1)}$, $\mathbf{G}^{(t)} \leftarrow \mathbf{G}^{(t-1)}$;
\State \Call{ITERATIVE-FUNCTION-1}{$\mathcal{D}$, $t$, $\mathbf{G}^{(t)}$, $\mathbf{\Lambda}^{(t)}$, $\mathcal{P}^{(t)}$, $T$}
\Until{$t == N_{iter}-1$}
%\Until{$T == T_{min}$}
\State \textbf{FINAL ITERATION:}
\State Set $t \leftarrow N_{iter}$; Set $T \leftarrow T_{min}$;
\State Set $\mathcal{P}^{(t)} \leftarrow \mathcal{P}^{(t-1)}$, $\mathbf{\Lambda}^{(t)} \leftarrow \mathbf{\Lambda}^{(t-1)}$, $\mathbf{G}^{(t)} \leftarrow \mathbf{G}^{(t-1)}$;
\State \Call{ITERATIVE-FUNCTION-2}{$\mathcal{D}$, $t$, $\mathbf{G}^{(t)}$, $\mathbf{\Lambda}^{(t)}$, $\mathcal{P}^{(t)}$, $T$}\end{algorithmic}
\end{algorithm}

\begin{algorithm}
\begin{algorithmic}[1]
\Function{ITERATIVE-FUNCTION-1}{$\mathcal{D}$, $t$, $\mathbf{G}^{(t)}$, $\mathbf{\Lambda}^{(t)}$, $\mathcal{P}^{(t)}$, $T$}
\State \textbf{Step 1:} Representation
\State For DBSOM, $\mathcal{P}^{(t)}$ is fixed;
%For DBSOM algorithm, $\mathcal{P}^{(t)}$ is kept fixed;
\State For ADBSOM, $\mathbf{\Lambda}^{(t)}$ and $\mathcal{P}^{(t)}$ are both fixed;
%For ADBSOM algorithm, $\mathbf{\Lambda}^{(t)}$ and $\mathcal{P}^{(t)}$ are kept fixed;
\State Compute $\mathbf{G}^{(t)}$ using equation (\ref{prot-1});
\State \textbf{Step 2:} Weighting (only ADBSOM algorithm)
\State Compute $\mathbf{\Lambda}^{(t)}$ using the suitable equation from Eq. (\ref{W-Glo-1V-Prod}) to Eq.(\ref{W-Loc-2C-Prod});
\State \textbf{Step 3:} Assignment
\State For DBSOM, $\mathbf{G}^{(t)}$ is fixed;
%For DBSOM algorithms, $\mathbf{G}^{(t)}$ is kept fixed;
\State For ADBSOM, $\textbf{G}^{(t)}$ and $\mathbf{\Lambda}^{(t)}$ are both fixed;
%For ADBSOM algorithm, $\textbf{G}^{(t)}$ and $\mathbf{\Lambda}^{(t)}$ are kept fixed;
\For{$1 \leq i \leq N$}
\State Let $w = f^{(t)}(y_i)$
%Let $v = f^{(t)}(y_i)$
\State DBSOM: let $z = \argmin_{1 \leq m \leq M} d^T(\mathbf{y}_i, \mathbf{g}^{(t)}_m)$
%DBSOM: let $x = \argmin_{1 \leq m \leq M} d^T(\mathbf{y}_i, \mathbf{g}^{(t)}_m)$
\State ADBSOM: let $z = \argmin_{1 \leq m \leq M} d_{\mathbf{\Lambda}}^T(\mathbf{y}_i, \mathbf{g}^{(t)}_m)$
%ADBSOM: let $x = \argmin_{1 \leq m \leq M} d_{\mathbf{\Lambda}}^T(\mathbf{y}_i, \mathbf{g}^{(t)}_m)$
\If{$w \neq z$}
\State $f^{(t)}(\textbf{y}_i) = z$;
\EndIf
\EndFor
\State \Return $\mathbf{G}^{(t)}$, $\mathbf{\Lambda}^{(t)}$, $\mathcal{P}^{(t)}$
\EndFunction
\end{algorithmic}
\end{algorithm}

\begin{algorithm}
\begin{algorithmic}[1]
\Function{ITERATIVE-FUNCTION-2}{$\mathcal{D}$, $t$, $\mathbf{G}^{(t)}$, $\mathbf{\Lambda}^{(t)}$, $\mathcal{P}^{(t)}$, $T$}
\Repeat
\State $test \leftarrow 0$;
\State \textbf{Step 1:} Representation
\State For DBSOM, $\mathcal{P}^{(t)}$ is fixed;
%For DBSOM algorithm, $\mathcal{P}^{(t)}$ is kept fixed;
\State For ADBSOM, $\mathbf{\Lambda}^{(t)}$ and $\mathcal{P}^{(t)}$ are both fixed;
%For ADBSOM algorithm, $\mathbf{\Lambda}^{(t)}$ and $\mathcal{P}^{(t)}$ are kept fixed;
\State Compute $\mathbf{G}^{(t)}$ using equation (\ref{prot-1});
\State \textbf{Step 2:} Weighting (only ADBSOM algorithm)
\State Compute $\mathbf{\Lambda}^{(t)}$ using the suitable equation from Eq. (\ref{W-Glo-1V-Prod}) to Eq.(\ref{W-Loc-2C-Prod});
\State \textbf{Step 3:} Assignment
\State For DBSOM, $\mathbf{G}^{(t)}$ is fixed;
%For DBSOM algorithms, $\mathbf{G}^{(t)}$ is kept fixed;
\State For ADBSOM, $\textbf{G}^{(t)}$ and $\mathbf{\Lambda}^{(t)}$ are both fixed;
%For ADBSOM algorithm, $\textbf{G}^{(t)}$ and $\mathbf{\Lambda}^{(t)}$ are kept fixed;
\For{$1 \leq k \leq N$}
\State Let $w = f^{(t)}(y_k)$
%Let $v = f^{(t)}(y_k)$
\State DBSOM: let $z = \argmin_{1 \leq m \leq M} d^T(\mathbf{y}_i, \mathbf{g}^{(t)}_m)$
%DBSOM: let $x = \argmin_{1 \leq m \leq M} d^T(\mathbf{y}_i, \mathbf{g}^{(t)}_m)$
\State ADBSOM: let $z = \argmin_{1 \leq m \leq M} d_{\mathbf{\Lambda}}^T(\mathbf{y}_i, \mathbf{g}^{(t)}_M)$
%ADBSOM: let $x = \argmin_{1 \leq m \leq M} d_{\mathbf{\Lambda}}^T(\mathbf{y}_i, \mathbf{g}^{(t)}_M)$
\If{$w \neq z$}
\State $f^{(t)}(\textbf{y}_i) = z$; $test \leftarrow 1$;
\EndIf
\EndFor
\Until{test == 0}
\State \Return $\mathbf{G}^{(t)}$, $\mathbf{\Lambda}^{(t)}$, $\mathcal{P}^{(t)}$
\EndFunction
\end{algorithmic}
\end{algorithm}

		\section{Applications} \label{SEC_apply}
		 In this section, we present an application of the proposed algorithms using 
		 %two synthetic datasets and  
		 two real-world data-sets.
%         The synthetic datasets will be used for showing the performance of the SOM algorithms under two different configurations of the components of two distributional variables. 
         These real world datasets are used for observing the algorithms' performance whit labeled data and unlabeled ones.
All the applications have been executed using the HistDAWass package in R \citep{HistDAWass}.

\subsection{Initialization and choice of the parameters}
Before launching a SOM algorithm some choices have to be done. In this paper, we do not propose a strategy for the selection of the best parameters of the SOM. In the literature \cite{Vesanto99}, some rule of thumbs is proposed for the SOM initialization according to some well-known side-effects (for example, the propensity of SOM to push all the objects in the corners of the map because of the kernel weighting impact). Hereafter, we list the choices for the parameters and the topologies chosen before launching the SOM algorithm.

We choice a 2D  hexagonal map of $5\sqrt{N}$ neurons \cite{Vesanto99}, where $N$ is the number of input objects. The map is rectangular having ratio (namely, the horizontal side of the map is longer than the vertical one), in order to let the map choice its direction of variability. This, in general, mitigates a SOM side-effect consisting in assigning objects into the Voronoi set of the BMU's associated with the corners of the map. Such an effect is more evident when using a dynamic clustering approach. Indeed, the whole SOM algorithm is based on the minimization of the cost function in Eq. \eqref{crit-1} or Eq. \eqref{crit-1-1}, and these functions are considered also for the assignment of data to the BMU's. To mitigate this effect, we propose to use toroidal maps \cite{Mount2011}. In the following, we present both planar and toroidal map results.

A second set of choices is related to the learning function and to the kernel one. First of all,  we use a Gaussian kernel function:
\begin{equation}\label{kernel_gau}
K^{(t)}(r,m)=\exp{\left(-\frac{d^2(r,m)}{2\cdot T(t)}\right)}
\end{equation}
where, $d^2(r,m)$ is the squared Euclidean distance in the topological space of the neurons between vertices (clusters) $r$ and $m$, $t$ is the generic epoch and $T$ is the kernel width (radius) at the epoch $t$.
Once defined the kernel, it is important choose the number of epochs for learning the map, the initial and the final kernel width (radius), and the rate of decreasing of the width (linear or exponential).
In the literature \cite{MATLAB_SOM}, for the batch SOM algorithm, it is suggested that the number of epochs for training the map  is lesser or equal than $N_{iter}=50$.

Fixed an initial value of the kernel width $T_{Max}=\sigma(1)$ and ending value with $T_{min}=\sigma(N_{iter})$, we use a power series decreasing function for the width of kernel as follows:
\begin{equation}\label{eq_learning_function}
T(t)=T_{Max}\left(\frac{T_{min}}{T_{Max}}\right)^\frac{t}{N_{iter}}.
\end{equation}
About the initial and final value of the kernel width, some heuristics proposed in the literature are mainly related to the classical SOM assignment (namely, the one performed using the distance between objects and BMU without considering the kernel). In \cite{Vesanto99}, it is suggested a value for $T_{Max}$ equal to $1/4$ the diameter of the map, decreasing along the epochs until it reaches 1. In our case, since the assignment is done consistently with the cost function, we experienced that the choices suggested by the literature lead to maps too folded and with a high topographic error.
We used a new heuristic as follows. Considering that the kernel function in Eq. \eqref{kernel_gau} ranges in $[0,1]$, we fix $T_{Max}$ such that two neurons having a distance equal to the radius of the map, namely, the half of the maximum topological distance between two neurons,  have a kernel value equal to $0.1$, and we fix $T_{min}$ such that two neighboring neurons have a kernel value equal to $0.01$. Denoting the diameter of the map in the topological space with $d_{Max}$, namely, the largest topological distance between two neurons of the map, the initial and the final value of $T$, considering Eq. \eqref{kernel_gau}, are determined as follows:
\begin{equation}\label{eq:tvalues}
T_{Max}=\sqrt{-\frac{\left(0.5\cdot d_{Max}\right)^2}{2\cdot \log{0.1}}}\;\;;\;\;T_{min}=\sqrt{-\frac{1}{2\cdot \log{0.01}}}.
\end{equation}
SOM is initialized randomly 20 times and the map with the lower final cost function is considered as the best run.

Considering that the proposed SOM algorithms as particular clustering algorithms, we evaluate the output results using internal and external validity indexes (except for unlabeled data).

		\subsection*{Internal validity indexes}
        For validating the map results, we consider the topographic error measure of the map, the Silhouette Coefficient as a base validity index for comparing the different algorithms and the Quality of Partition index developed in Ref. \citep{IrpinoESWA}. Further, we propose some extensions of the silhouette index for SOM.

A classical validity index for SOM is the topographic error \citep{topo_err_96} $\mathcal{E}_T$. It is a measure of topology preservation and it is computed as follows: for all input vectors, the respective
best and second-best matching units are determined. An error is counted if best and second-best matching units of an input vector are neighbors on the map. The total error is the ratio of the sum of the error counts with respect to the cordiality of the input vectors. It ranges from 0 to 1, where 0 means perfect topology preservation. It is obtained as follows:
\begin{equation}\label{eq:topoerr}
\mathcal{E}_T = \frac{1}{N}\sum\limits_{i = 1}^N {u(i)}.
\end{equation}
where
$$u(i) = \left\{ {\begin{array}{*{20}{l}}
{1,}&{\mathrm{best-\;and\;second-BMU\;not - adjacent}}\\
{0,}&\mathrm{otherwise}
\end{array}} \right.$$
        We recall that the Silhouette Coefficient was defined for clustering algorithms and it corresponds to the average of the silhouette scores $s(i)$ computed for each input data vector. In particular, $s(i)$ \citep{Silh_87} is obtained as:
        $$ s(i)=\frac{b(i)-a(i)}{max[a(i),b(i)]}$$
        where $a(i)$ is the average distance between the input vector $i$ and the ones assigned to the same cluster (say A, the cluster of $i$), while $b(i)$ is the minimum among the average distances computed from the $i$ to the input vectors assigned to each cluster except the one to which $i$ belongs (say B, the second best cluster of $i$) . The Silhouette Coefficient is as follows:
        \begin{equation}\label{eq:sil}
        \mathcal{S}=\frac{1}{N}\sum\limits_{i=1}^N s(i).
        \end{equation}
        The Silhouette Coefficient ($\mathcal{S}$) ranges from $-1$ to $1$, where $1$ represents the best clustering result since each cluster is compact and well separated from the others. Even if in its original formulation it has a complexity of $\mathcal{O}(N^2)$  for a generic distance, in the case of (squared) Euclidean distances it is possible to show that it is only $\mathcal{O}(N)$ \footnote{See the appendix for the proof.}. For reducing the computational time of the index, a simplified version of the Silhouette Coefficient was also proposed by \cite{Silh_Campello06}, where the distances are computed with respect to the prototypes of the cluster only. We use also this index denoting it with $\mathcal{S}_C$.

        Regarding also the SOM solution and the considerations about the topographic error, we remark that the obtained SOM map contains neurons that are represented by prototypes that are generally similar if the neurons are adjacent. This could reduce the Silhouette Coefficient even if the map is a good representation of the cluster  structure of the input data. To adapt the Silhouette index to the SOM, we propose to  mix together  the advantages of  the Silhouette Coefficient in revealing a good cluster structure and the ones related to the topographic error. We thus propose to modify the $b(i)$ calculation considering it as the second-best cluster of $i$ as the  cluster of input vectors that is not adjacent to the neuron of the $i$ BMU. The Silhouette score of each $i$ is calculated as above, but $b(i)$ is obtained without considering those neurons adjacent to A (namely, the BMU of $i$). We propose the same strategy for the $\mathcal{S}_C$ index and we denote the two new coefficients with $\mathcal{S}_\mathcal{E}$ and $\mathcal{S}_{C\mathcal{E}}$ respectively. We remark that in the case of ADBSOM the distance used for the computation of $a(i)$ and $b(i)$ are the adaptive distances used in the algorithm.

%        The quality of partition index (QPI) is a generalization of the classic R-square statistics. It is the ratio of the Between Sum of Squared deviation and the Total Sum of Squared deviation of the clustered dataset. Using the statistics developed for distributional data in \citep{IrpVer2015}, the QPI has been adapted to asses the quality of clusterings of distributional data with adaptive distances in \citep{IrpinoESWA}. The QPI is defined as follows:
%         \begin{equation}\label{eq:QPI}
%         QPI= 1-\frac{WSS}{TSS},
%         \end{equation}
%         where $WSS$ is the value of the criterion in Eq. (\ref{crit_1}) or (\ref{crit_2}) at the convergence of the algorithm, and $TSS$ is the total sum of squares of the data.

%         \textcolor{red}{I PROPOSE TO EXPLICITLY WRITE IN THE PAPER THE FORMULA OF $tss$}

\subsection*{External Validity Indexes}
Let $N$ be the number of instance of a data table, $\mathcal{P}=\{C_1,C_2,\ldots,C_M\}$ the clusters obtained by a clustering algorithm and $\mathcal{P'}=\{C'_1,C'_2,\ldots,C'_K\}$ the classes of the labeled instances. Generally, $M$ is required to be equal to $K$, but in our case we can also consider $M\neq K$.
Let $C_m$ (resp. $C'_k$) the instances of cluster $m$ (resp. of class $k$), and $a_m=|C_m|$ (resp., $b_k=|C'_k|$) its cardinality. Let $n_{mk}=|C_m \cap C'_k|$
the number of instances of cluster $m$ being in class $k$.

For evaluating the results of the algorithms, for dataset with labeled instances, we use three external validity indexes: the \emph{Adjusted Rand Index} (ARI) \cite{Hubert1985}, the \emph{purity} (Pur) \cite{Manning_2008}  and the \emph{Normalized Mutual Information} ($NMI$) \cite{Meila2007}.

The ARI index \cite{Hubert1985} is widely used for assessing the concordance between apriori partition and the partition provides by the algorithm: the index varies between $-1$ and $1$ where the more the index approaches $1$ the more the two partitions are similar. The ARI shows the ability of the algorithm to recover the original classification.
\begin{equation}\label{eq:ARI}
ARI=\frac{ \sum_{mk}{\binom{n_{mk}}{2}} - [ \sum_{m} {\binom{a_{m}}{2} } \sum_{k} {\binom{b_{k}}{2} } ] / {\binom{N}{2} } } { \frac{1}{2} [ \sum_{m} { \binom{a_{m}}{2} } + \sum_{k} { \binom{b_{k}}{2} } ] - [\sum_{m} { \binom{a_{m}}{2} } \sum_{k} { \binom{b_{k}}{2} } ] / {\binom{N}{2} } }.
\end{equation}

The Normalized Mutual Information (NMI)
index between apriori partition and the partition provided by the algorithm, is computed as follows:
\begin{equation}\label{eq:NMI} NMI=\frac{I(\mathcal{P},\mathcal{P'})}{|H(\mathcal{P})+H(\mathcal{P'})|/2}\end{equation}
where $I$ is the mutual information
$$I(\mathcal{P},\mathcal{P'})=\sum_{m}\sum_{k}\frac{n_{mk}}{N}log \frac{n_{mk}}{a_{m}b_{k}}$$
$H$ are the entropies:
$$H(I(\mathcal{P}))=-\sum_{m}\frac{a_m}{N}log\frac{a_m}{N}\;\mathrm{and} \;H(I(\mathcal{P'}))=-\sum_{k}\frac{b_k}{N}log\frac{b_k}{N}.$$
ARI and NMI is maximal when the number of classes is equal to the number of clusters. This can be problematic when evaluating the SOM results. In fact, the considering the SOM as a clustering algorithm it is frequent that the number of clusters is greater than the number of apriori classes. We propose to consider these measures together with the \emph{purity} index, namely, another external validity index that considers also this possibility.

Purity index ($pur$) measures the homogeneity of clusters with respect to apriori partition. The index is calculated as follows:
\begin{equation}\label{eq:purit}pur=\frac{1}{N}\sum_{m}\argmax_k(n_{mk})\end{equation}
It consists in evaluating the fraction of labeled
instances of the majority class in each cluster for all the clustering. It varies in $[0; 1]$, where $1$ indicates that all clusters are pure, namely, they contain only labeled instances of one class.
However, \emph{pur} presents a major drawback. It over
estimates the quality of a clustering having a large number of clusters that is the typical situation of a SOM output partition.

Thus, we propose to read together the two sets of external indexes following these guidelines: if a SOM has both a higher value of \emph{pur} and of NMI and ARI, this means that the number of non-empty clusters (namely, neurons that are BMU's for at least one instance) are close to the number of apriori classes, while, if NMI and ARI are relatively low and \emph{pur} index is high it means that the apriori class labeled instances are shared among a set of neurons identifying pure clusters. Obviously, if NMI, ARI and \emph{pur} are low the resulting SOM is less able to recognize the apriori class structure.

In the remainder of the applications, in the tables we denote respectively the algorithms with \textit{St.}, the classic BSOM algorithm with each variable standardized using the standard deviation for distributional variables,  \textit{P1}, the ADBSOM algorithm with the automatic detection of the relevance weights on each variable of the whole dataset, \textit{P2}, the ADBSOM algorithm with the automatic detection of relevance weights for the components of each variable included in the dataset, \textit{P3}, the ADBSOM algorithm with the automatic detection of relevance weights for each variable and each neuron, and \textit{P4}, the ADBSOM algorithm with relevance weights automatically detected for the components of each variable and each neuron.

\subsection{Real-world datasets}
A first dataset comes from an experiment on activity recognition of people doing \textit{Daily and Sports Activities} that is publicly available from the UCI repository \cite{UCI}. In particular, the raw data consists in the triaxial gyroscope and accelerometers measurements of five sensors (two on the arms and on the legs, and one on the thorax) of eight people performing 19 different activities for 5 minutes \citep{altun_comparative_2010}. Each 5 minutes session of activity is represented by 60 5-seconds time windows described by the histograms of the 125 measurements recorded in that time window. In this case, considering that the records are labeled according to the person and the activity,  we show how the proposed maps are able to represent the different activities or people (for a specific activity) using some external validity indexes.

The second dataset describes the population pyramids of 228 countries of the world observed in 2014. Considering that a population pyramid is the description of the age distribution for the male and the female component of a population, the dataset is described by only two distributional variables: the male age distributional variable and the female age one. We will refer to this dataset naming it as the ``AGE PYRAMIDS" dataset.
The AGE PYRAMIDS dataset does not contain indications about the cluster structure. In this case, to compare the algorithms we will use some internal validity measures like the silhouette index \citep{Silh_87}, the topographic error of the map \citep{topo_err_96} and the quality of partition index proposed in Ref. \citep{CarvalhoBS16}.

\subsection{Human Behavior Recognition dataset}\label{realdata3}
       The dataset that we consider here can be downloaded from the University of California Irvine machine learning repository\footnote{http://archive.ics.uci.edu/ml/}. It collects data on 19 activities performed by 8 different people performed in sessions of 5 minutes. Table \ref{tab: persons} shows the description of the people involved and table \ref{tab: activities} the list of activities monitored.

\begin{table}[htbp]
\centering
\caption{Description of the 8 individuals in the experiment.}
\label{tab: persons}
\begin{tabular}{|l|llll|}
\hline
ID & gender & age & height & weight  \\ \hline
1  & F      & 25  & 170    & 63      \\
2  & F      & 20  & 162    & 54      \\
3  & M      & 30  & 185    & 78      \\
4  & M      & 25  & 182    & 78      \\
5  & M      & 26  & 183    & 77       \\
6  & F      & 23  & 165    & 50      \\
7  & F      & 21  & 167    & 57      \\
8  & M      & 24  & 175    & 75      \\ \hline
\end{tabular}
\end{table}

\begin{table}[htbp]
\centering
\caption{List of activities in the data set}
\label{tab: activities}
\begin{tabular}{ p{\dimexpr 0.05\linewidth-2\tabcolsep}
                   p{\dimexpr 0.45\linewidth-2\tabcolsep}
                   p{\dimexpr 0.05\linewidth-2\tabcolsep}
                   p{\dimexpr 0.45\linewidth-2\tabcolsep} }
 \hline
 $\#$&Action&$\#$&action\\
\hline
1  & sitting   &11 & walking on a treadmill at 4km/h in 15º inclined position                                                            \\
2  & standing  &12 & running on a treadmill at 8km/h     \\
3  & lying on the back &13 & exercising on a stepper     \\
4  & lying on the right side &14 & exercising on a cross trainer            \\
5  & ascending stairs  &15 & cycling on exercise bike in horizontal position  \\
6  & descending stairs &16 & cycling on exercise bike in vertical position   \\
7  & standing still in an elevator &17 & rowing   \\
8  & moving around in an elevator &18 & jumping  \\
9  & walking in a parking lot  &19 & playing basketball  \\
10 & walking on a treadmill at 4 km/h in flat position & & \\ \hline
\end{tabular}
\end{table}

The research group that collected the data set has extensively used it to compare classification algorithms \citep{altun_comparative_2010} and classification software packages \citep{barshan_recognizing_2014}, study inter-subject and inter-activity variability  \citep{barshan_investigating_2016}.

The data is collected by means of five MTx3-DOF units, manufactured by Xsens Technologies. Sensor units are calibrated to acquire data at a sampling frequency of 25Hz. Each unit has a tri-axial accelerometer, a tri-axial gyroscope, and a tri-axial magnetometer. Sensor units are placed on the arms, the legs and the thorax of the subject's body.
We do not consider the magnetometer sensors. The reason is that the magnetometer recordings reflect the direction of the activity with respect to the Earth's magnetic North and this information somehow contaminates the data set.
Thus, the data set has 30 continuous variables (5 units x 2 sensors x 3 axes). For each one of the 30 time series recorded at a sampling frequency of 25Hz during 5 minutes was broken into 60 (no-overlapping) time-windows of 5 seconds each of them containing 125 measures. Each set of 125 measures is aggregated into an equi-depth (or equiprobable) histogram where each bin contains the $10\%$ of the observed values.

As a result of the aggregation, for each activity and person, we have 30 (histogram) variables with 60 histograms. As a result, we have a histogram-data table of 9120 rows (60 time windows $\times$ 8 people $\times$ 19 activities), where each row is a window of 5 seconds of a given person performing one of the activities.

\subsection*{Walking on a treadmill at 4 km/h in flat position}
Using the activity recognition dataset, we selected one of the activities and we analyze if it is possible to recover a class structure in data. In particular, we want to test our algorithm about its ability at discriminating people doing the same activity. Among the 19 activities, we did a preliminary exploration and we noted that the activity $\#10$ \textit{Walking on a treadmill at 4 km/h in flat position} shows a particular differentiation among people. In this case, the subset is a table of $14,400$ histograms having $480$ rows, namely, $60$ 5-seconds-windows for each one of the $8$ people, and $30$ columns, namely, the measurements of the $5$ tri-axial sensors recording acceleration (accelerometers) and angular speeds (gyroscope).

\begin{figure}[htbp]
			\centering\caption{Activity Recognition dataset: PCA on people while walking on a treadmill at 4 km/h in flat position.}\label{Fig_PCA_Walking}
			\includegraphics[width=0.7\textwidth]{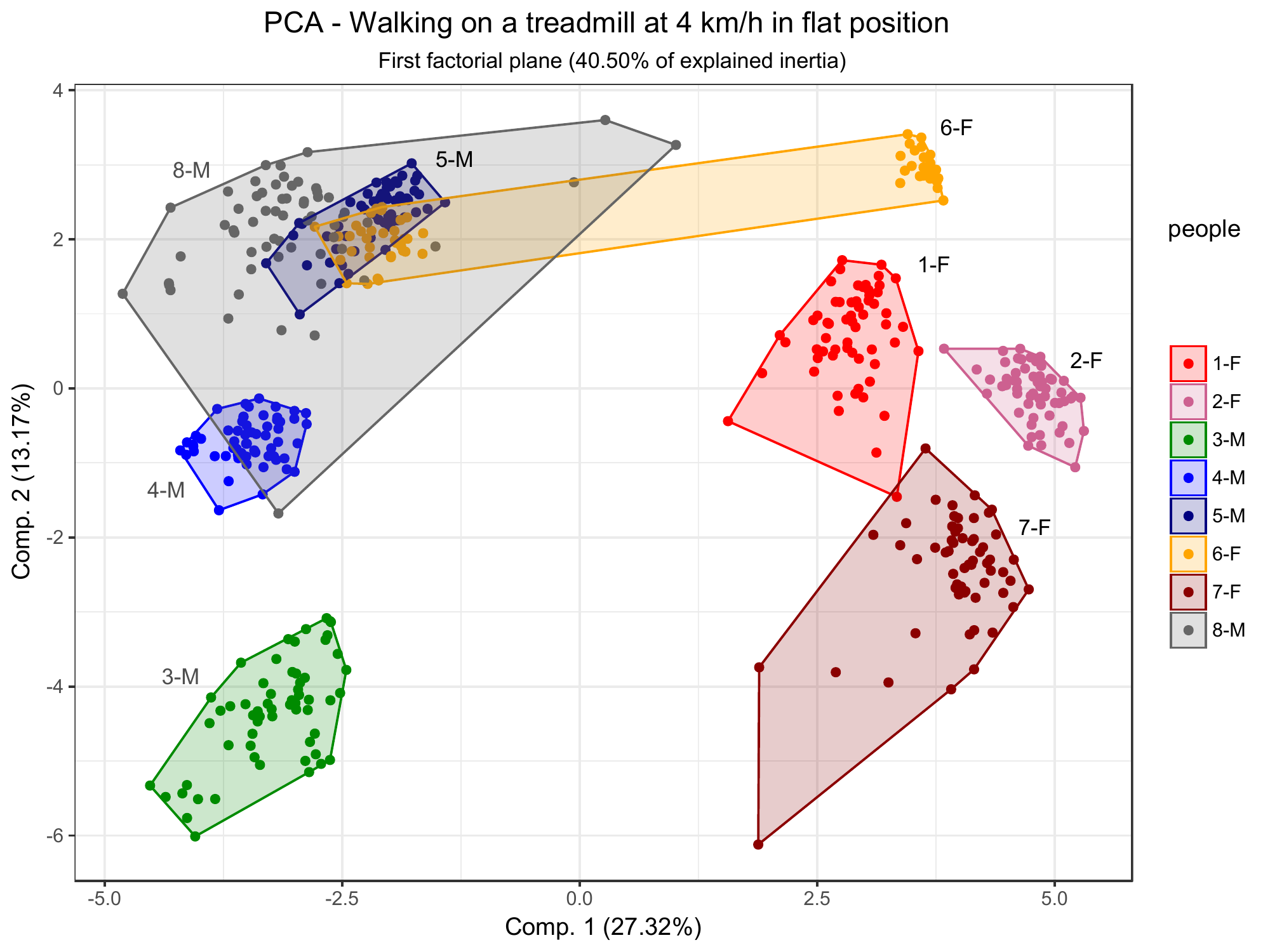}
		\end{figure}

In Fig. \ref{Fig_PCA_Walking}, we show the plot of individuals, namely, the time windows recorded for all the people, in the first factorial plane obtained by a PCA performed using the method proposed in \cite{VERDE_PCA_2017}. The first factorial plane explains the $40.50\%$ of the total inertia and the 60 time-windows of each person are contained in a convex polygon that is colored differently for each person. Since the variables are $30$, the percentage of inertia explained by the first factorial plane is not so high. However, we notice that the eight people appear quite separated and, in particular, we can see that females are on the right of the plane, while males are on the left except for person 6 (a woman that presents a particular pattern showing two different ways of performing the activity during the five-minutes session).

        In this case, we run the five SOM algorithms initializing two types of maps: a planar and a toroidal one. The map is a $16\time 8$ hexagonal grid. The size of sides of the map has been chosen such that the cardinality of the neurons is close to the $5\sqrt{480} \approx110$ and each side has an even number of neurons (this is required for the toroidal map using a hexagonal grid).  The main validity indexes are reported in Tab. \ref{TAB_Val_WALKING}. The external validity indexes assume that the reference labels are the ones identifying the people.
        %\begin{table}[!h]
%\centering
%\resizebox{\textwidth}{!}{
%\begin{tabular}{llrrrrrrrrr}
%  \hline
%  & & \multicolumn{6}{l}{Internal validity}&\multicolumn{3}{l}{External validity}\\
%Map & Met. & $\mathcal{S}$ &  $\mathcal{S}_\mathcal{E}$ &  $\mathcal{S}_{C}$ & $\mathcal{S}_{C\mathcal{E}}$ & QPI & $\mathcal{E}_T$ & ARI & NMI & Pur \\
%  \hline
%   7x7 & St. & .3747 & .5116 & .5770 & .7200 & .8013 & .1562 & .7144 & .8236 & .9479 \\
%    7x7 & P1 & .3554 & .6375 & .5789 & .8098 & .8950 & .1146 & .7317 & .8489 & 1.000 \\
%    7x7 & P2 & .4705 & .6502 & .6416 & .7918 & .8787 & .0500 & .6970 & .8258 & .9479 \\
%    7x7 & P3 & .5088 & .7460 & .6603 & .8562 & .9505 & .1375 & .7675 & .8661 & 1.000 \\
%    7x7 & P4 & .4555 & .6654 & .6992 & .8080 & .9391 & .2479 & .7989 & .8726 & .9437 \\
% \hline
%  10x10 & St.& .1861 & .3850 & .4760 & .6511 & .8030 & .1562 & .5425 & .7584 & .9271 \\
%  10x10 & P1 & .2777 & .3545 & .5794 & .6606 & .7742 & .1625 & .3436 & .6755 & .6771 \\
%  10x10 & P2 & .4141 & .4667 & .6748 & .7232 & .7133 & .2146 & .4291 & .7186 & .6979 \\
%  10x10 & P3 & .3155 & .5097 & .6190 & .7397 & .9624 & .0854 & .7585 & .8411 & 1.000 \\
%  10x10 & P4 & .3664 & .6504 & .6347 & .8028 & .9719 & .0958 & .6031 & .7726 & .9417 \\
%   \hline
%\end{tabular}
%}
%\caption{Activity recognition dataset, walking on a treadmill at 4 km/h in flat position: validity indexes. ARI is computed for data labeled by the people.}\label{TAB_Val_WALKING}
%\end{table}

\begin{table}[ht]
\centering
\resizebox{\textwidth}{!}{
\begin{tabular}{lrrrrrrrr}
  \multicolumn{9}{c}{$16\times 8$ hexagonal \textsc{planar} map}\B \\
  \hline
   & \multicolumn{5}{l}{Internal validity indexes}&\multicolumn{3}{l}{External validity indexes}\\
 Met. & $\mathcal{S}$ &  $\mathcal{S}_\mathcal{E}$ &  $\mathcal{S}_{C}$ & $\mathcal{S}_{C\mathcal{E}}$ &  $\mathcal{E}_T$ & ARI & NMI & Pur \\
  \hline
 St. & 0.3065 & 0.4306 & 0.5914 & 0.6888 & 0.1812 & 0.4872 & 0.7391 & 1.0000 \\
   P1 & 0.3588 & 0.4769 & 0.6132 & 0.7224 & 0.0750 & 0.5748 & 0.7630 & 1.0000 \\
   P2 & 0.3354 & 0.5181 & 0.5823 & 0.7380 & 0.2167 & 0.5646 & 0.7650 & 1.0000 \\
   P3 & 0.5519 & 0.6889 & 0.7300 & 0.8358 & 0.1292 & 0.7054 & 0.8305 & 0.9958 \\
   P4 & 0.2810 & 0.5384 & 0.5658 & 0.7546 & 0.0521 & 0.5894 & 0.7974 & 1.0000 \\
    \hline
            \multicolumn{9}{c}{$16\times 8$ hexagonal \textsc{toroidal} map} \T \B\\
    \hline
     & \multicolumn{5}{l}{Internal validity indexes}&\multicolumn{3}{l}{External validity indexes}\\
 Met. & $\mathcal{S}$ &  $\mathcal{S}_\mathcal{E}$ &  $\mathcal{S}_{C}$ & $\mathcal{S}_{C\mathcal{E}}$ &  $\mathcal{E}_T$ & ARI & NMI & Pur \\
  \hline
   St. & 0.2888 & 0.5886 & 0.5221 & 0.7644 & 0.0229 & 0.4528 & 0.7636 & 1.0000 \\
   P1 & 0.3903 & 0.7308 & 0.6045 & 0.8594 & 0.0021 & 0.4979 & 0.7798 & 1.0000 \\
   P2 & 0.3954 & 0.6818 & 0.5981 & 0.8180 & 0.0062 & 0.6077 & 0.8194 & 1.0000 \\
   P3 & 0.3587 & 0.7154 & 0.5958 & 0.8538 & 0.0062 & 0.5567 & 0.7918 & 0.9958 \\
   P4 & 0.4043 & 0.7564 & 0.5823 & 0.8633 & 0.0021 & 0.5322 & 0.8063 & 1.0000 \\
   \hline
\end{tabular}
}
\caption{Activity recognition dataset, walking on a treadmill at 4 km/h in flat position: validity indexes. External validity indexes assume people as labels.}\label{TAB_Val_WALKING}
\end{table}

In Tab. \ref{TAB_Val_WALKING}, results show that toroidal SOMs have lower topographic errors and are more compact with respect to the planar maps. Looking at the external validity indexes, there are not great substantial differences. Normalized mutual information index appears slightly better for toroidal maps (except for the P3 algorithm), while the obtained non-empty Voronoi sets of each neuron are very pure.  The adjusted Rand indexes are not so high, but this is due to the different number of obtained non-empty Voronoi sets with respect to the number of classes.

Internal validity indexes also confirm better compactness of the clusters identified by the Voronoi sets associated with the BMUs for the toroidal map with respect to the planar one.

\begin{figure}[htbp]
 			\centering\caption{Activity Recognition dataset, walking on a treadmill at 4 km/h in flat position: P4 \textsc{toroidal} SOM count map.}\label{Fig_P4 H_WALK}
 			\includegraphics[width=0.9\textwidth]{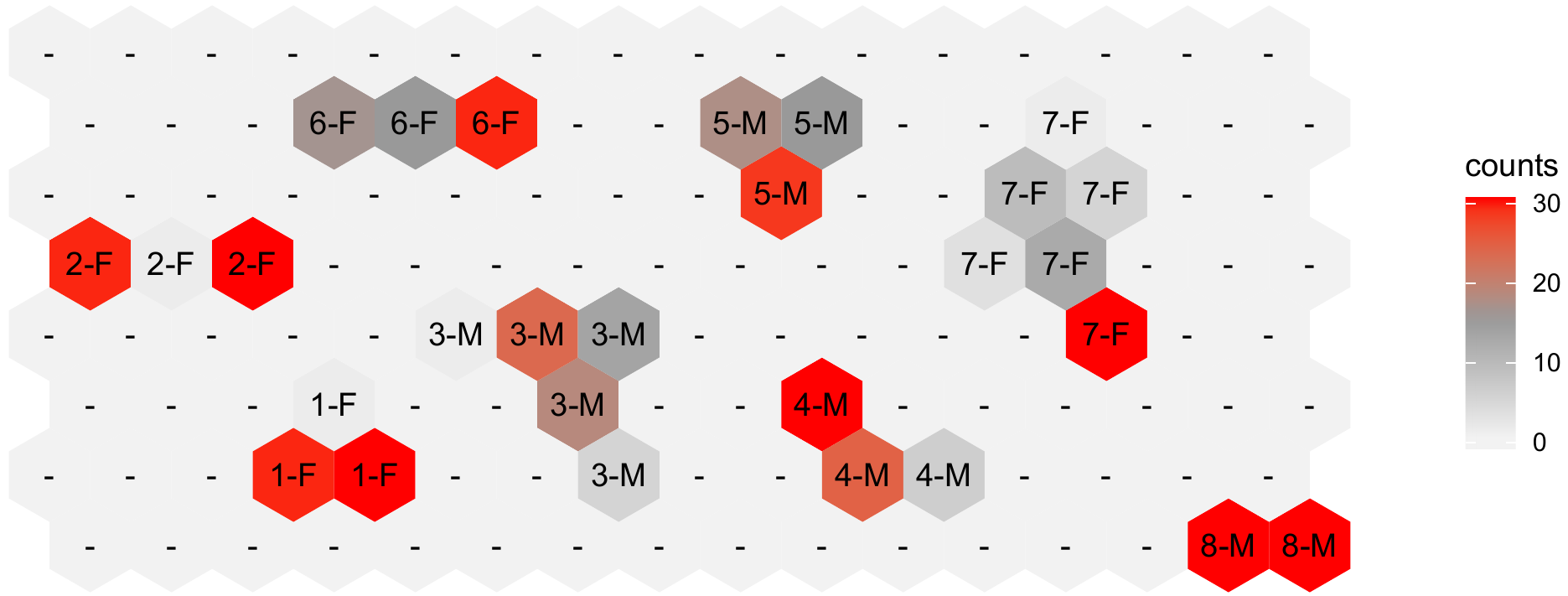}
\end{figure}

\begin{figure}[htbp]
 			\centering\caption{Activity Recognition dataset, walking on a treadmill at 4 km/h in flat position: P3 \textsc{planar} SOM count map.}\label{Fig_P3 pla_WALK}
 			\includegraphics[width=0.95\textwidth]{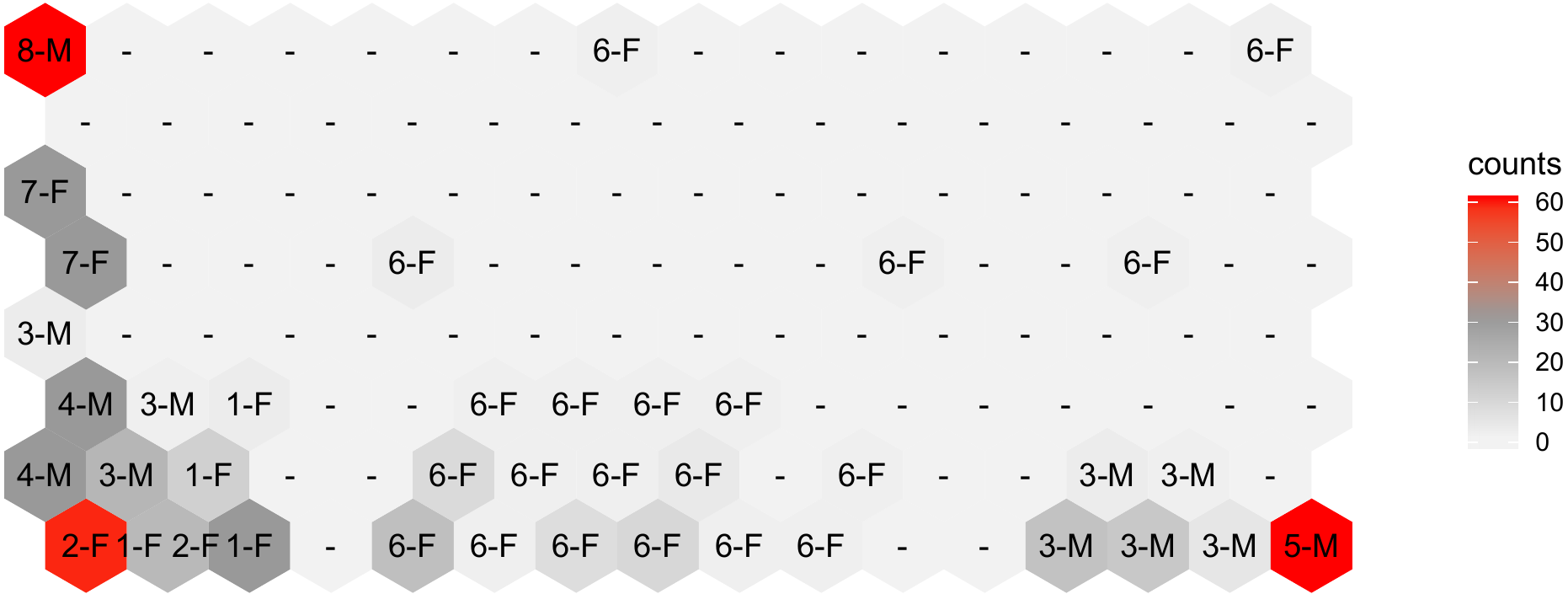}
\end{figure}

Fig. \ref{Fig_P4 H_WALK} and \ref{Fig_P3 pla_WALK} show the counts of the Voronoi sets (according to the intensity of the colors) of the neurons of the map obtained using P4 algorithm for the hexagonal map, namely, the map having the lower topographic error and best values of internal validity indexes, and the one obtained from the P3 algorithm and a planar map, namely, the one having the highest ARI and NMI index.
Each neuron is labeled according to the labels of the objects contained in its Voronoi set. We observe that the only neuron that has two labels is second on the bottom-left of the planar map.  Looking at the maps, we observe how the pushing-toward-the-edges effect is evident for the planar map, while, for the toroidal map, we may appreciate how the cluster of neurons are more evident and separated. This suggests that the topographic error, together with internal validity indexes could be good hints for deciding what map could be more explicative of the class structure of this kind of datasets.

\begin{figure}[htbp]
 			\centering
 			\includegraphics[width=0.9\textwidth]{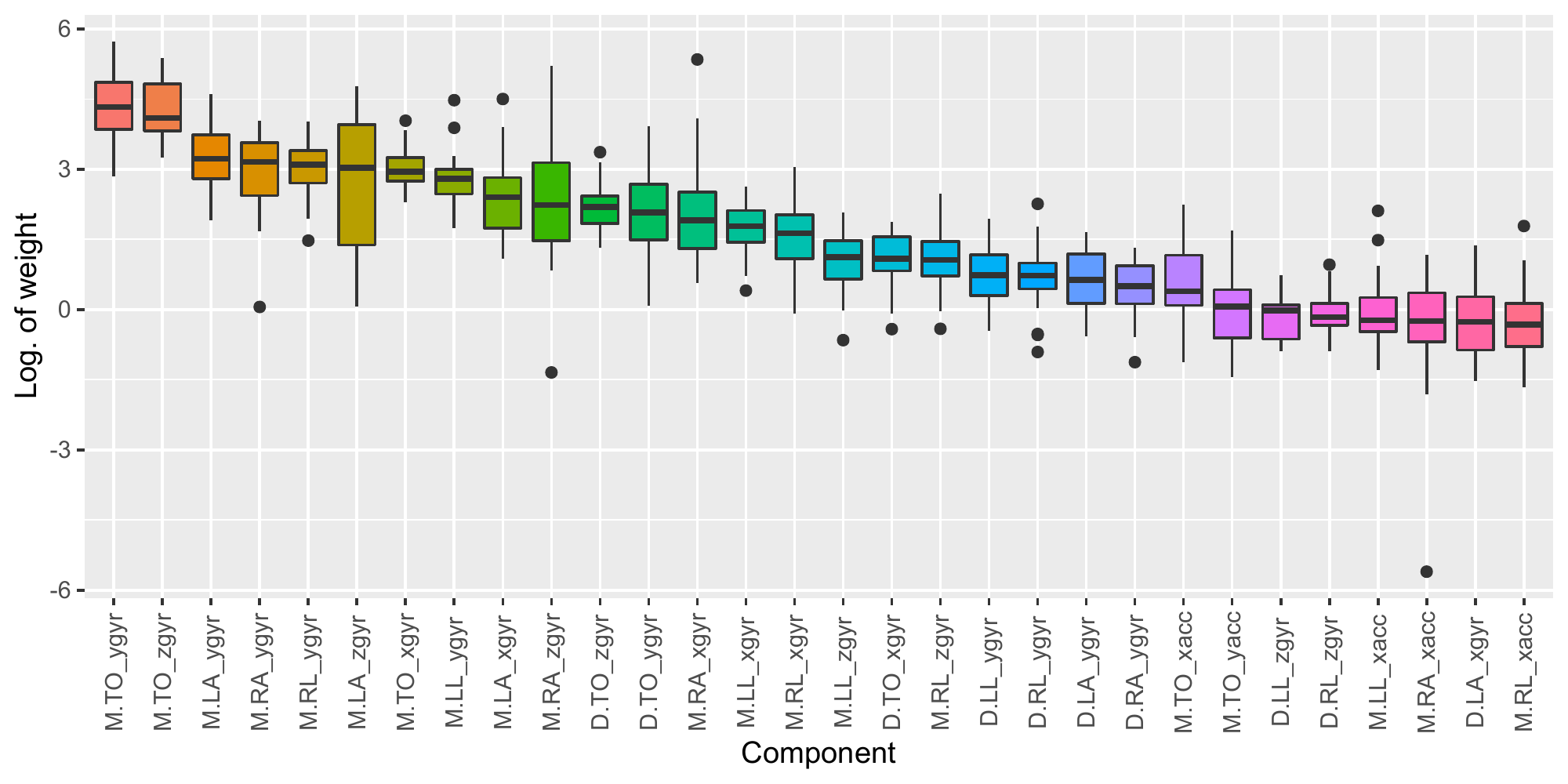}
\includegraphics[width=0.9\textwidth]{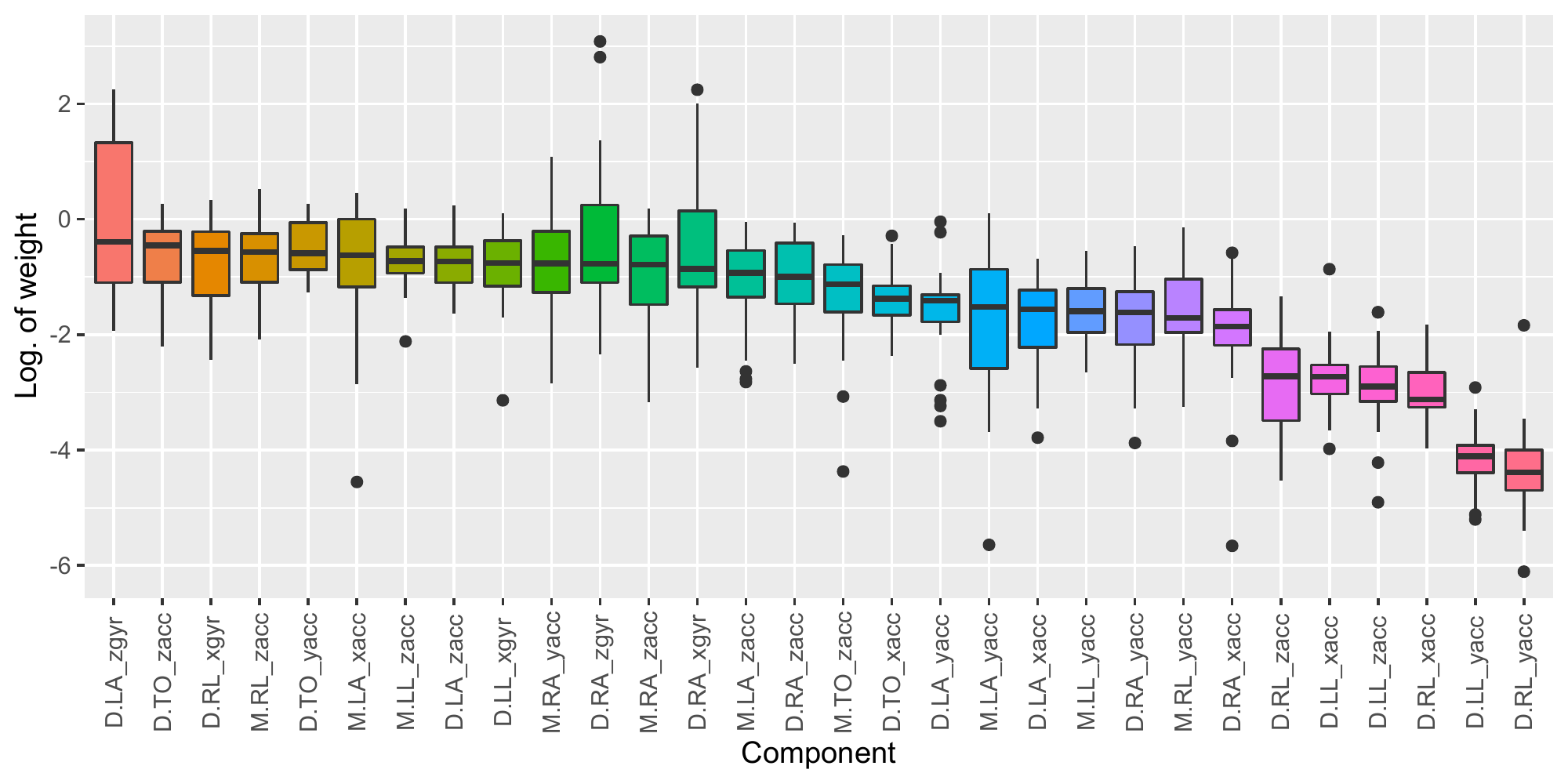}
 		\caption{Activity Recognition dataset, walking on a treadmill at 4 km/h in flat position: box-plot of the logarithms of weights for each component of variables sorted according to the median weights: M denotes average component, D the dispersion one, TO, RA, LA, RL and LL are the position of the sensors.}\label{Fig_BP_som_wei}
        \end{figure}

In the remainder of this paragraph, we continue to describe results for the P4 algorithm with a toroidal map\footnote{The authors can supply the R code and detailed results as supplementary material or on request.}.
Since the P4 algorithm assigns a relevance weight to each variable for each neuron, it is interesting to observe what variables are more relevant for SOM. In fig. \ref{Fig_BP_som_wei}, we show the box-plots of the logarithms of the relevance weights for the two maps. We remark that we used the box-plot since each variable may have a different weight for each neuron of the first map. For the sake of readability, we ordered the box-plot according to the median value observed for the relevance weights.

 Generally, it is interesting to note that the thorax gyroscope variables assume greater weights (the box-plots on the top left side of the plots), while lower values are generally assumed by accelerometer measures (on the left part).

\begin{figure}[htbp]
 			\centering
 \includegraphics[width=0.7\textwidth]{SOM_counts_P4_H.pdf}
 \includegraphics[width=0.75\textwidth]{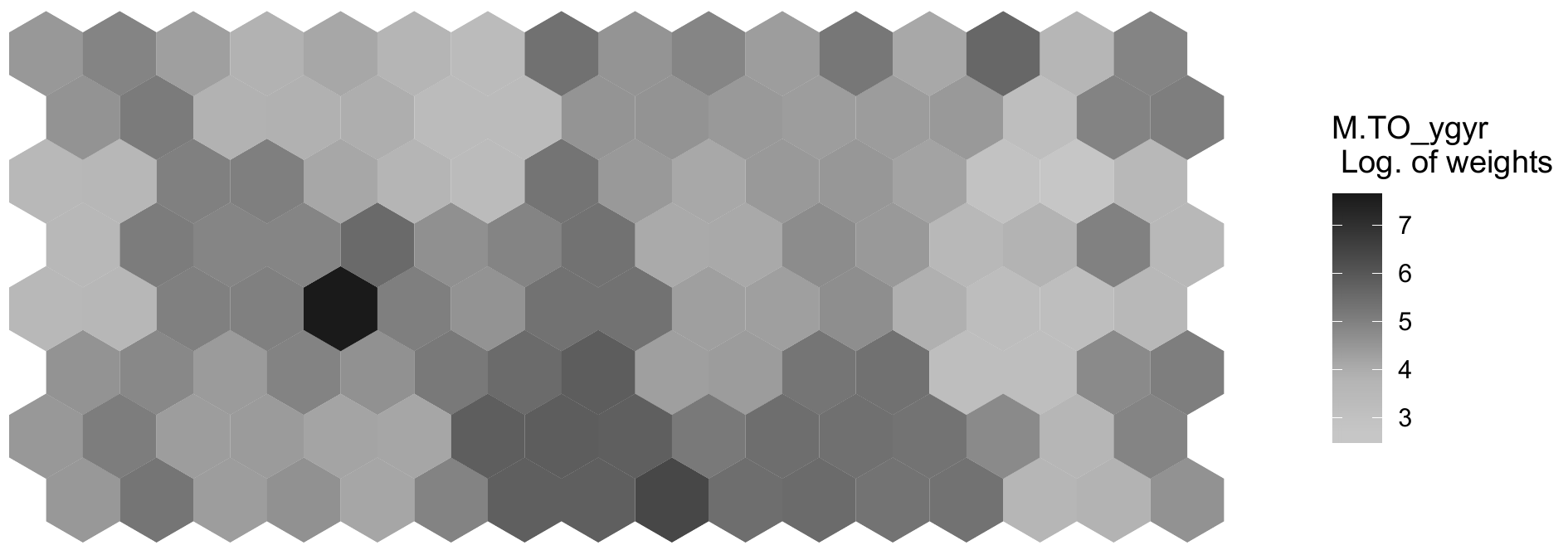}
 \includegraphics[width=0.75\textwidth]{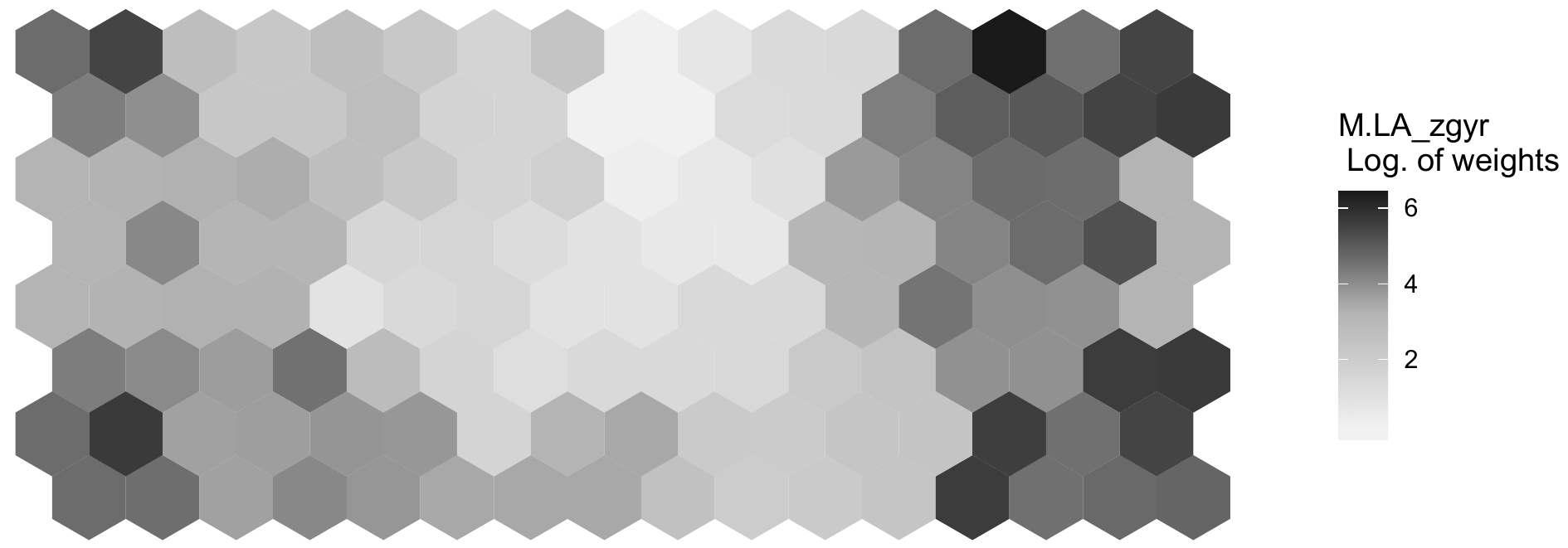}
            \caption{Activity Recognition dataset, walking on a treadmill at 4 km/h in flat position, results of the P4 algorithm using a toroidal map. On the top the count map. At the center, the map showing the logarithms of the relevance weights for the component having the highest median relevance. At the bottom, the map showing the logarithms of the relevance weights for the component having the highest variability of the relevance weights among all the neurons.}\label{Fig_som_wei_map}
 		\end{figure}

It is interesting to observe on the map what are the neurons where components assume greater relevance with respect the others. In Fig. \ref{Fig_som_wei_map}, are reported the maps of the relevance weights for two (of 60) components, namely the average component of the gyroscope measurements on the $y$ axis of the sensor positioned on the torax ($M.TO\_ygyr$) and the average component of the gyroscope measurements on the $z$ axis of the sensor positioned on the left arm ($M.LA\_ygyr$). The choice is motivated by the fact that, looking at Fig. \ref{Fig_BP_som_wei}, $M.TO\_ygyr$ component is the one having the highest median relevance weights among neurons and $M.LA\_ygyr$ has the highest variability. To allow the reader a more immediate reading of the results, the count map of Fig. \ref{Fig_P4 H_WALK} is replicated at the top of Fig. \ref{Fig_som_wei_map}.
It is worth noting that, $M.TO\_ygyr$, that is related to the torsion of the thorax from left to right, is more relevant for male than for female people this because the torso of a male differs from the one of a female and this impact on the rotational change on the $y$ axes. Looking at the map for  $M.LA\_ygyr$, the relevance is higher for people 1, 2, 7 and 8, namely three females and one male. This is related to the forward and backward movement of the left arm while walking and the variability in the importance of this component within people may be caused, for example, by their handedness. Indeed, it ranges from positive (namely, highly relevant) log values to negative (namely, lowly relevant) ones.
Other interpretations could be done observing the other variables but, for the sake of brevity, we don't go ahead, but we confirmed that the use of relevance weights may enrich the interpretation of the results.
%%%%%%%%%%%%%%%%%%%%%%%%%%%%%%%%%%%%%%%%%%%%%%%%%%%%%%%%%%%%%%%%%%%%%%%%%%
\subsection{World countries population pyramids dataset}\label{realdata2}
		The second application evaluates the effectiveness of the proposed algorithms on a dataset recording the population age-sex pyramids of 228 World countries. The dataset is provided by the Census Bureau of USA in 2014 and it is included in the \texttt{HistDAWass}\footnote{\texttt{https://cran.r-project.org/package=HistDAWass}} package developed in R.
		
		The input dataset consists in the description of 228 countries according to the age relative frequency distributions for the male and the female part of the population, namely, it is a table of $228\times2$ histograms.

The demographic evolution of a population is usually represented by three prototypical pyramids: constrictive, expansive and stationary. These represent the main stages of the demographic evolution of a population and are often used as an indicator of life quality in a country.
         
\begin{figure}[htbp]
			\centering\caption{Types of population pyramid}\label{Fig_types_pyr}
			\includegraphics[width=0.8\textwidth]{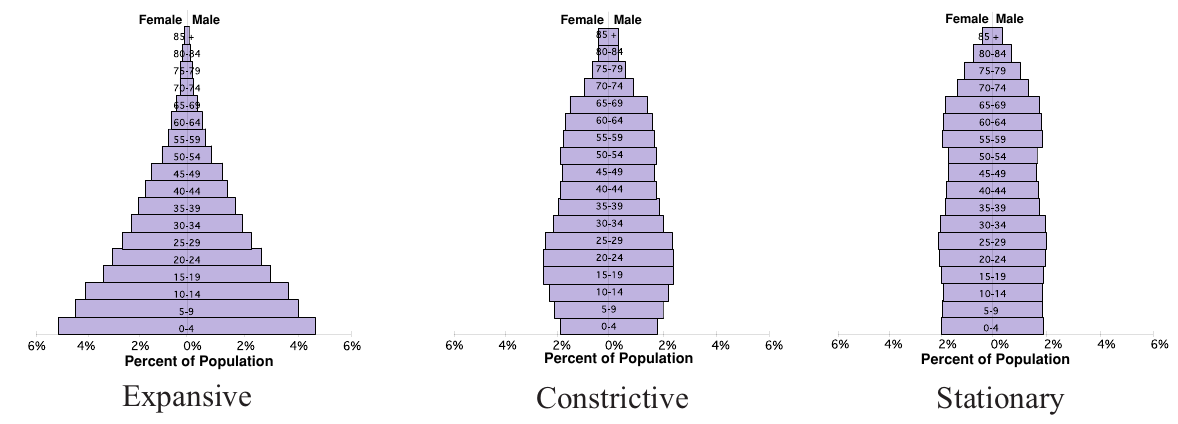}
		\end{figure}

         In Fig. \ref{Fig_types_pyr}, we report the three prototypical pyramid structures \cite[Ch. 5]{Atlas}.
% * <antonio.irpino@gmail.com> 2018-04-25T15:29:34.828Z:
%
% see here for proxy meaning https://dictionary.cambridge.org/dictionary/english/proxy#dataset-british
%
% ^.

We remark that data is not labeled according to these three prototypical models. Using BSOM we want to see if such prototypical situations arise in the data and how well the map is able to represent the data structure and the variables importance in the map generation.

%%%%%%%%%%%%%%%%%%%%%%%%%%%%%

        Also in this case, we run the five SOM algorithms initializing two types of maps: a planar and a toroidal one. The map is a $16\time 8$ hexagonal grid. The size of sides of the map has been chosen such that the cardinality of the neurons is close to the $5\sqrt{228} \approx110$ and each side has an even number of neurons (this is required for the toroidal map using a hexagonal grid).  
        %The main validity indexes are reported in Tab. \ref{TAB_Val_WALKING}. The external validity indexes assume that the reference labels are the ones identifying the people.
        Two sized maps, a medium sized hexagonal map ($6\times6$) and a large sized one map ($8\times8$), has been chosen and has been randomly initialized 50 times.
        Table \ref{TAB_Val_AGE_Pyr} shows the validity indexes related to the final results of the proposed algorithms.

          \begin{table}[htbp]
\centering

\begin{tabular}{lrrrrr}
\multicolumn{6}{c}{$10\times 8$ hexagonal \textsc{planar} map}\B \\
  \hline
   & \multicolumn{5}{l}{Internal validity indexes}\\
 Met.& $\mathcal{S}$ &  $\mathcal{S}_\mathcal{E}$ &  $\mathcal{S}_{C}$ & $\mathcal{S}_{C\mathcal{E}}$ & $\mathcal{E}_T$\\
 St. & 0.3432 & 0.6134 & 0.6670 & 0.8262 & 0.2368 \\
   P1 & 0.2085 & 0.5186 & 0.6176 & 0.7949 & 0.1886 \\
   P2 & 0.2688 & 0.5634 & 0.6505 & 0.8181 & 0.1842 \\
   P3 & 0.3027 & 0.6060 & 0.6490 & 0.8283 & 0.1360 \\
   P4 & 0.2884 & 0.6419 & 0.6572 & 0.8459 & 0.0965 \\
    \hline
            \multicolumn{6}{c}{$10\times 8$ hexagonal \textsc{toroidal} map} \T \B\\
    \hline
     & \multicolumn{5}{l}{Internal validity indexes}\\
  St. & 0.2820 & 0.6340 & 0.6367 & 0.8371 & 0.1360 \\
   P1 & 0.3179 & 0.6252 & 0.6437 & 0.8257 & 0.1228 \\
   P2 & 0.3424 & 0.6889 & 0.6474 & 0.8580 & 0.1316 \\
   P3 & 0.2129 & 0.6307 & 0.5954 & 0.8312 & 0.0921 \\
   P4 & 0.3951 & 0.7417 & 0.6880 & 0.8841 & 0.0570 \\
  \hline
\end{tabular}
\caption{Age Pyramids dataset: validity indexes.}\label{TAB_Val_AGE_Pyr}
\end{table}

Considering the topographic errors, the use of adaptive distances provides better results than standard ones. Silhouette indexes $\mathcal{S}_\mathcal{E}$ and $\mathcal{S}_{C\mathcal{E}}$ also show an average better compactness and separation in the clustering structure induced by the SOM with adaptive distances.
In particular, for this dataset, P4 algorithm, namely, the one assigning weights to the components of each variable for each cluster, returns a lower topographic error and shows more compactness for the clusters defined by the Voronoi sets of the prototypes associated with the neurons.

\begin{figure}[htbp]
			\centering\caption{Age Pyramid dataset.P4 SOM algorithm: map of counts and the codes of countries in each Voronoi set of the map BMU's.}\label{Fig_Pyr_counts}
			\includegraphics[width=0.8\textwidth]{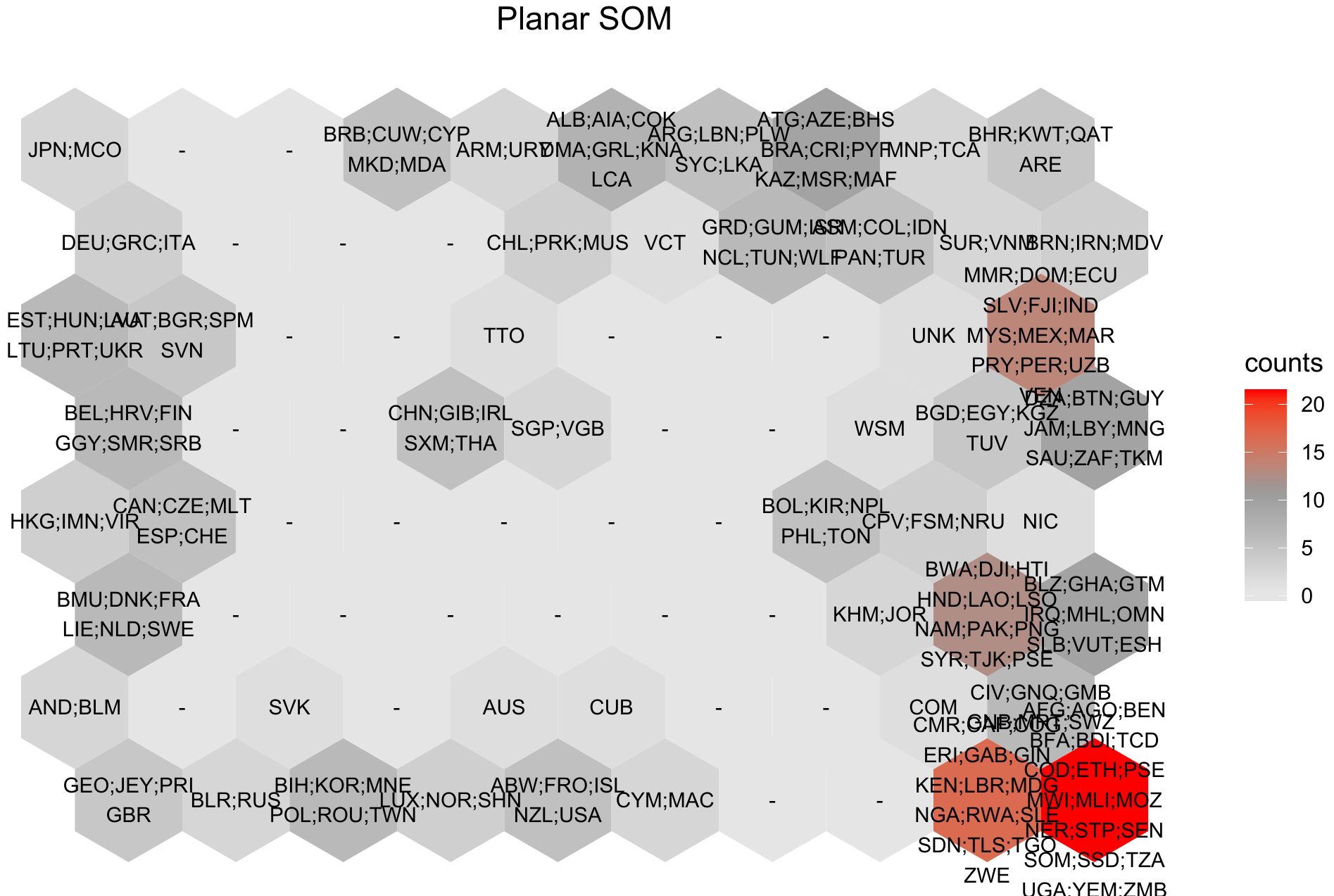}
            \includegraphics[width=0.8\textwidth]{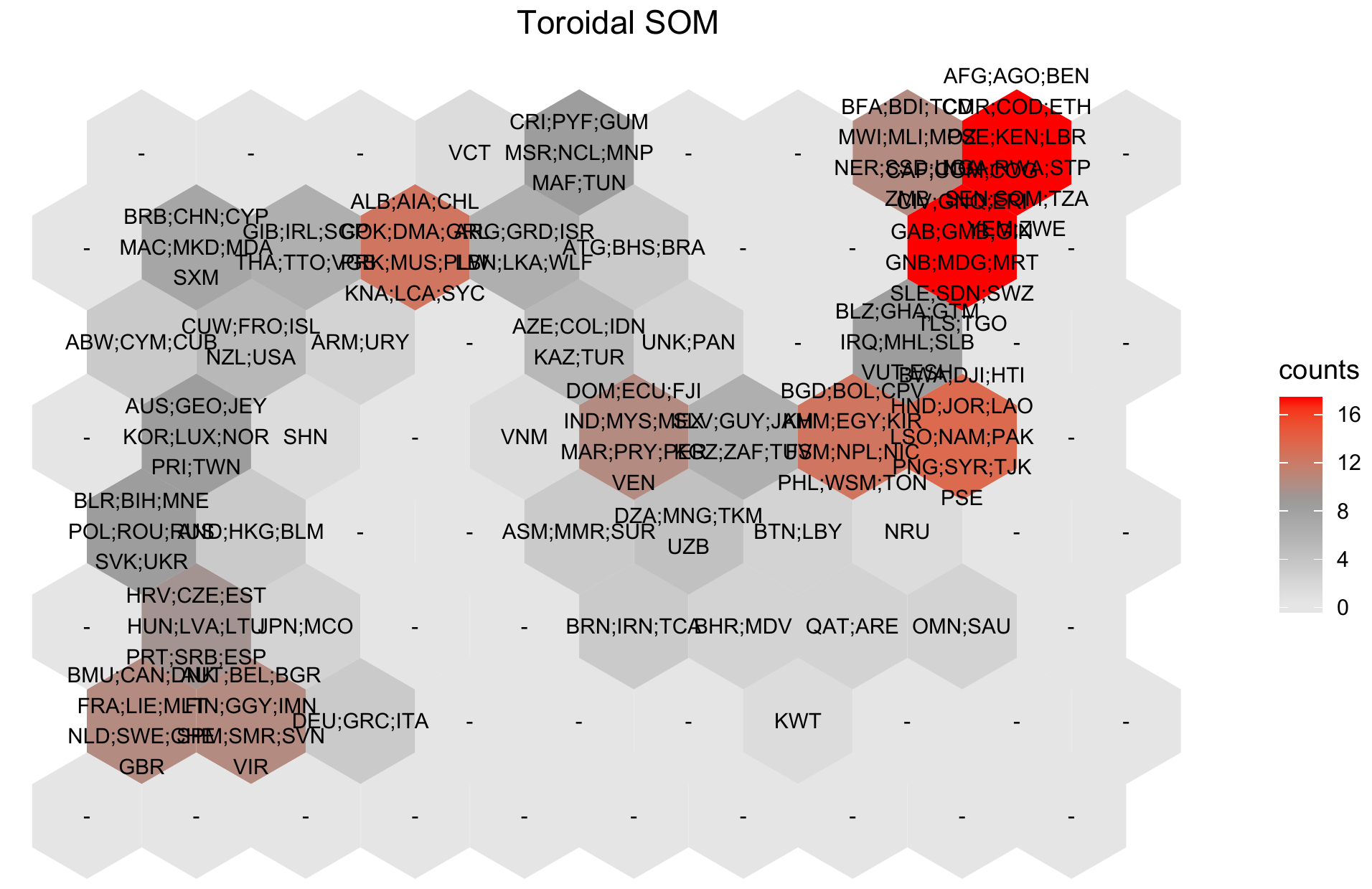}
		\end{figure}

\begin{figure}[htbp]
			\centering\caption{Age Pyramid dataset. Toroidal SOM using P4 algorithm: prototypes map.}\label{Fig_P4_tor_protos}
			\includegraphics[width=0.9\textwidth]{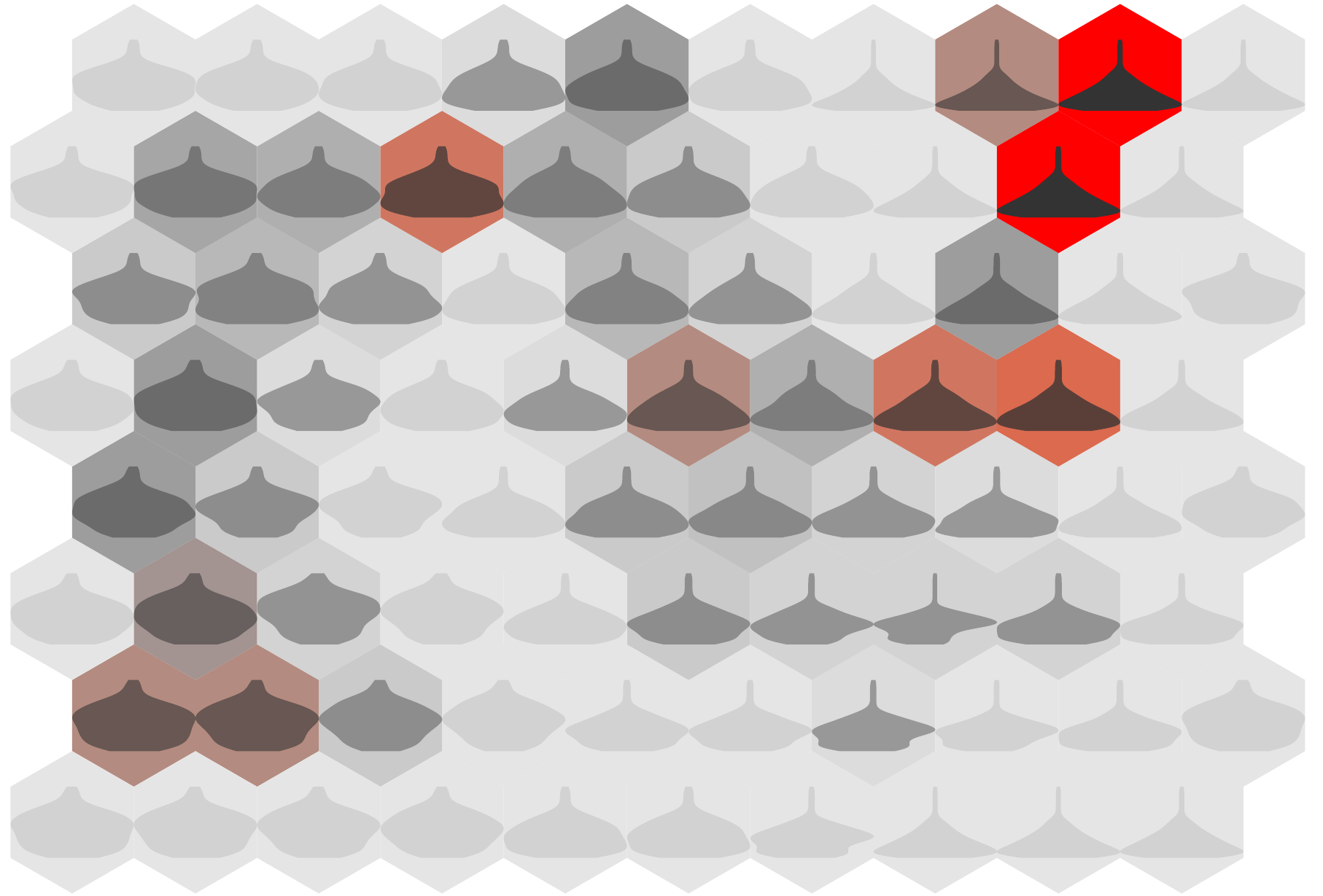}
		\end{figure}

Taking the best results in terms of topographic error, in Fig. \ref{Fig_Pyr_counts} are shown the maps with the counts and the ISO 3166-1 alpha-3 Country codes of data attracted by each BMU for the P4 algorithms both in the planar and in the toroidal map case.
As shown in Fig. \ref{Fig_Pyr_counts} planar map tends to push on the corners and the border of the map the data, leaving a central empty zone that is not well justified by the data. Indeed, in this dataset are described population structures of countries that are not so clustered in the reality, but the one which are at one of the stages described in Fig. \ref{Fig_types_pyr}. Thus, we consider the toroidal map description of the data more coherent than the planar one, showing a sort of path of transition from countries with an \emph{Expansive} type of population (namely, the ones in the left top corner) toward \emph{Constrictive} ones (namely, the ones in the mid of the path), and, finally, to \emph{Stationary} ones. This type of pattern is better visualized in Fig. \ref{Fig_P4_tor_protos}, where is represented the population pyramid (namely, a particular type of codebook map for data described by two distributional variables where each prototype is described by two juxtaposed smoothed frequency distributions) associated with the prototype of each BMU of the toroidal SOM.

\begin{figure}[htbp]
			\centering\caption{Age Pyramid dataset. Toroidal SOM using P4 algorithm: log of relevance weights plot.}\label{Fig Pyr_P4_wei}
			\includegraphics[width=0.45\textwidth]{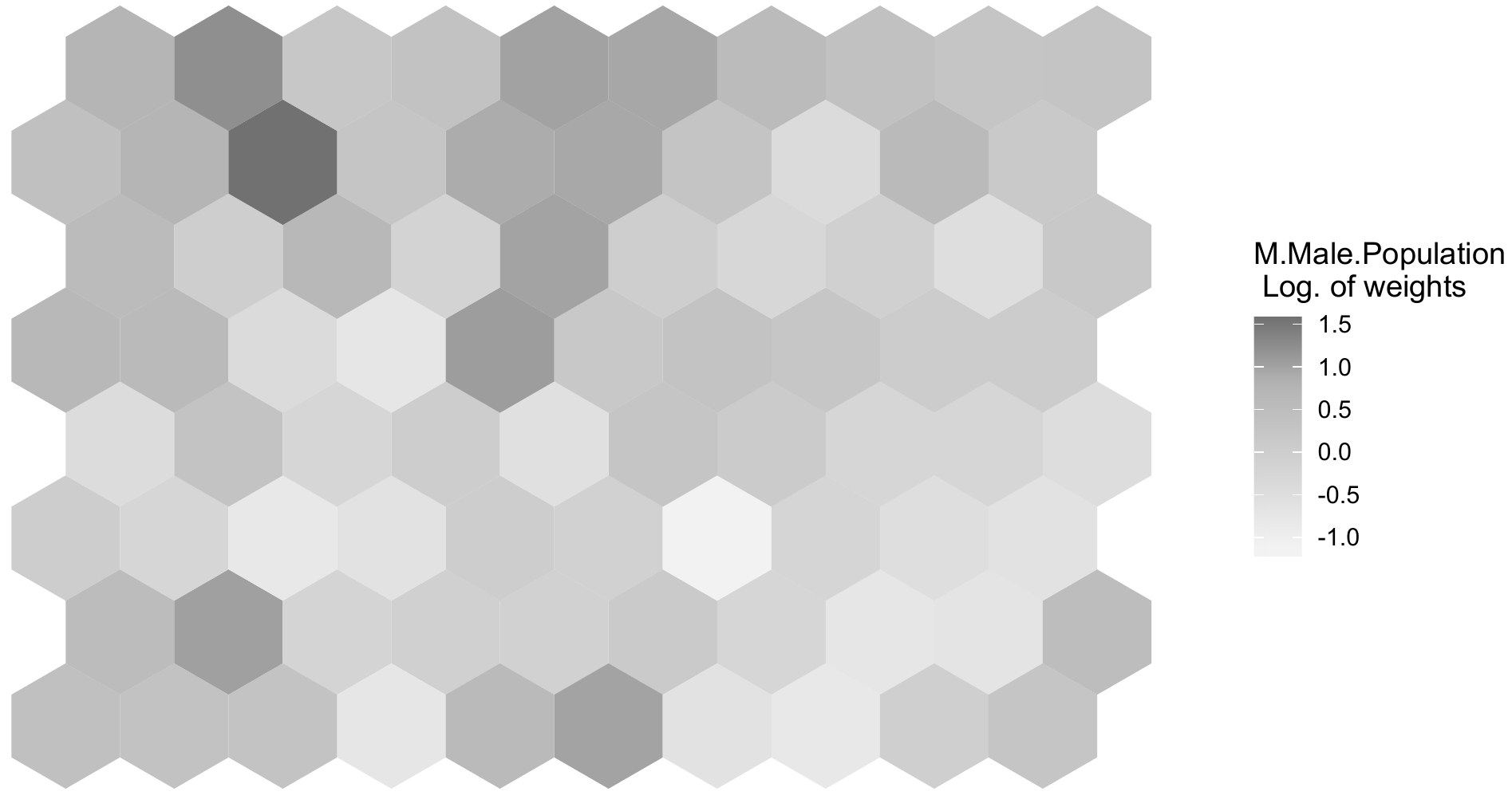}
\includegraphics[width=0.45\textwidth]{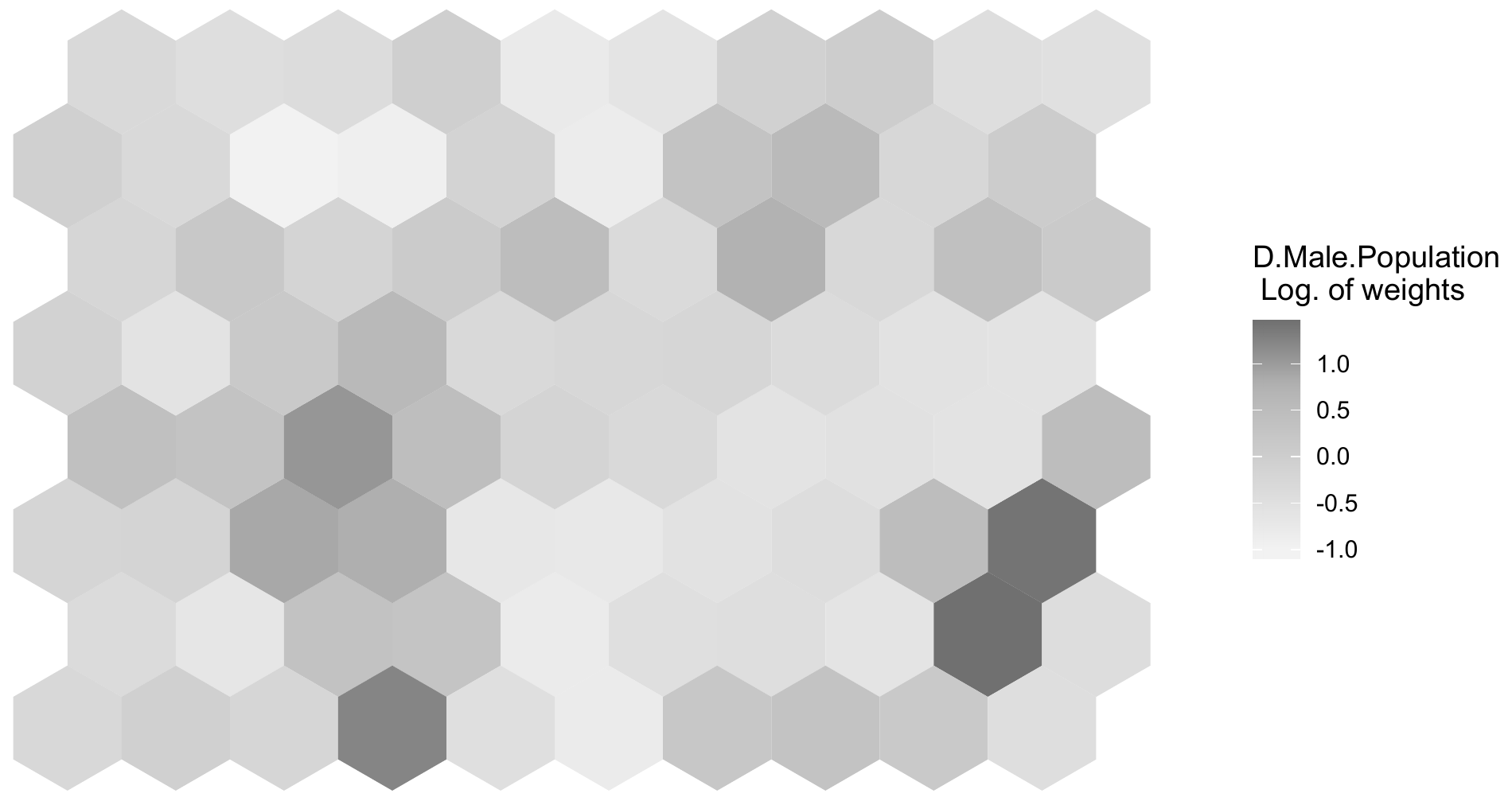}
\includegraphics[width=0.45\textwidth]{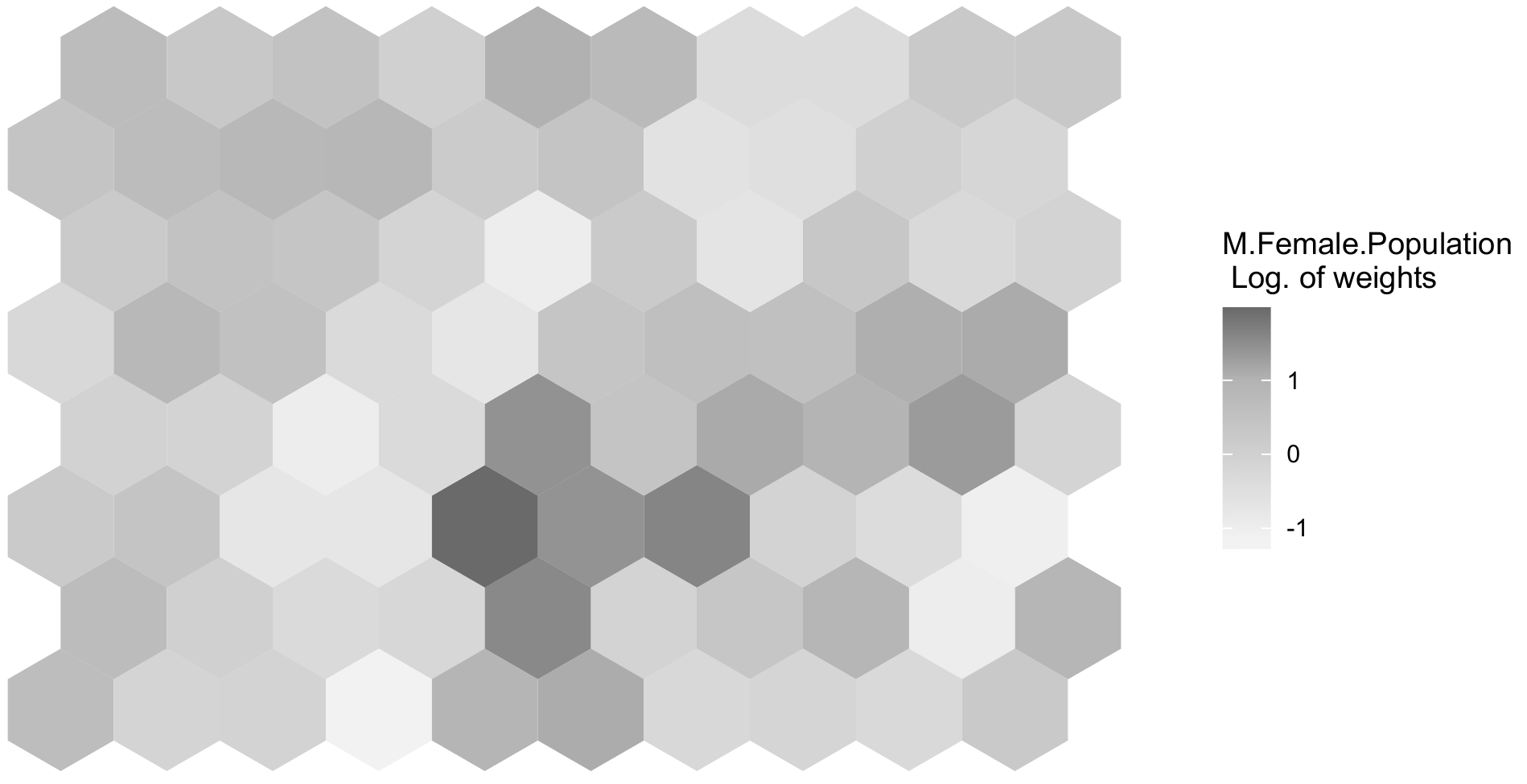}
\includegraphics[width=0.45\textwidth]{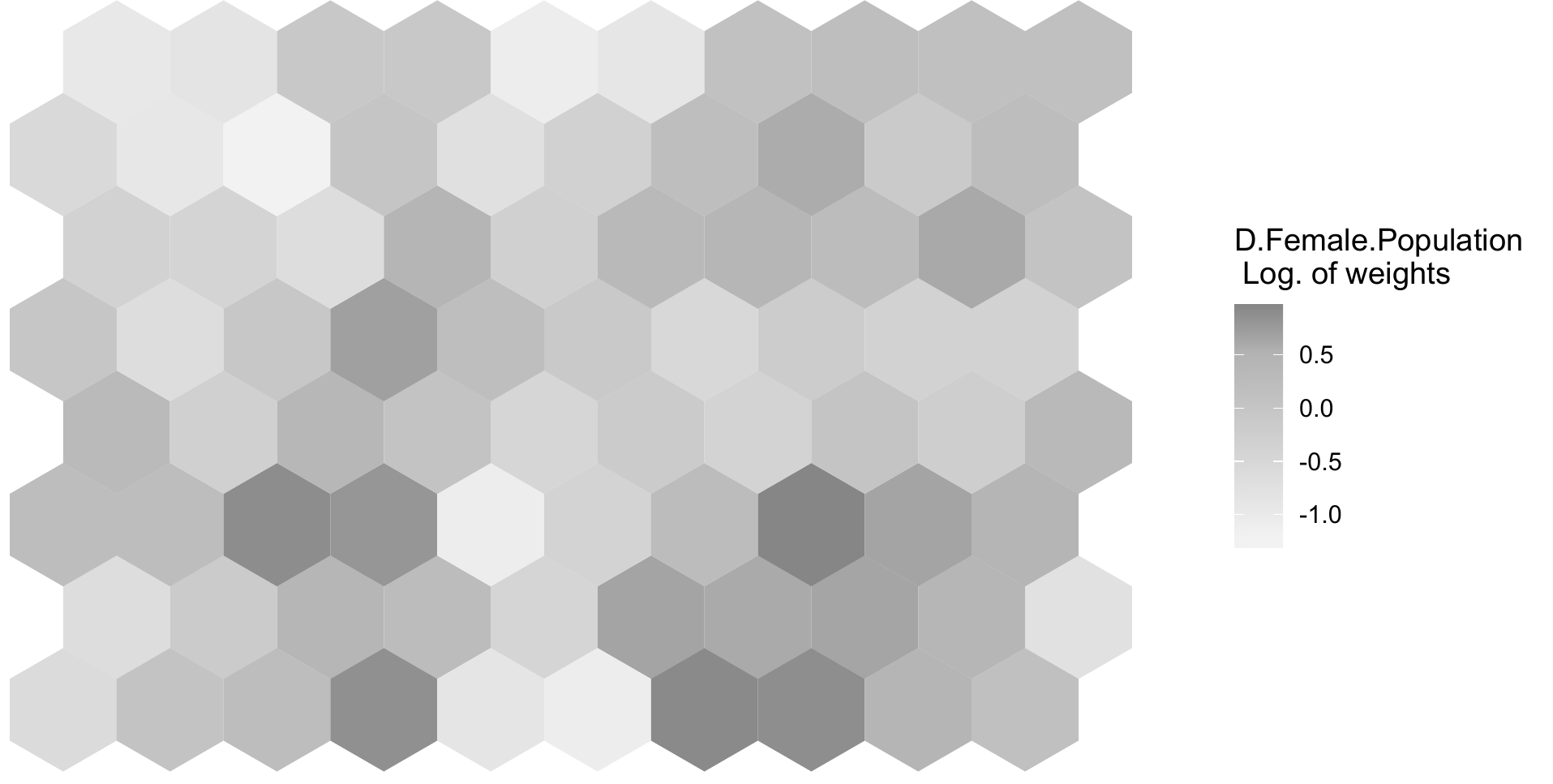}
		\end{figure}

In Fig. \ref{Fig_P4_tor_protos}, the three population structures are almost evident starting from the \emph{expansive} model on the top-left side of the map, passing by the \emph{constrictive} model at the center until the \emph{stationary} model on the bottom right. An interesting zone is the bottom zone of the map. Looking at the Fig.\ref{Fig_Pyr_counts}, we observe that the zone is a representation of the Arab states of the Persian Gulf where the population distributions present some particular patterns related to the story and the economy of the region. What is clear is that a path from the expansive to the constrictive and stationary model arise confirming the demographic theories that assume the models as three phases of an evolutionary path of a population.
Finally, in Fig. \ref{Fig Pyr_P4_wei} are shown the map of the logarithms of the relevance weights associated with each Voronoi set of the BMU of the neurons. It is interesting to note that, for example, for Persian gulf countries the variability components of the male and female population age distribution is more relevant, while, in general, the average component of the male age population is the most important in the cluster definition especially for the top-left zone of the map where there is the passage from a \emph{constrictive} to a \emph{stationary} shaped population.

 		\section{Conclusions}\label{SEC_conl}
The two Batch Self-Organizing Maps (SOMs) methods proposed DBSOM, to extend classical Batch SOM algorithm for distributional data, and ADBSOM, to innovative Batch SOM algorithms, using adaptive distances.
 DBSOM and ADBSOM training algorithms are based on the optimization of different objective functions. that is performed in alternates steps: two  for DBSOM and three  for ADBSOM.

In the representation step, DBSOM and ADBSOM algorithms compute the prototypes (vectors of distributions) of the clusters related to the neurons. They are achieved consistently with the optimal solution of the objective functions, having used the $L_2$ Wasserstein distance between pairs of distributions vector.

The main contribution to Classical batch SOM algorithm is to overcome the classical assumption of SOM that all the variables have the same relevance for training the SOM.
Indeed, it is well known that some variables are less relevant than others in the clustering process.

In particular, using the $L_2$ Wasserstein distance between 1D distributions, we have extended the use of SOM for data described by distributions and we have also introduced adaptive distances in the SOM algorithm according to four different strategies. The particular choice of adaptive distances, differently from other variable weighting schemes, does not require to tune further parameters.

The weighting step of ADBSOM calculates the relevance-weights of the distributional valued variables. This is achieved conforming to the optimal solution provided in the paper.
The squared $L_2$ Wasserstein distance between two distributions can be decomposed into two components: a squared Euclidean distance between the means and a $L_2$ Wasserstein distance between the centered quantile functions associated to the distributions. This second component allows taking more into account the variability and shape of the distributions. Relevance weights are automatically learned for each of the
measure components to emphasize the importance of the different characteristics (means, variability, and shape) of the distributions.
These weights of relevance are calculated for each cluster or for the entire partition such that their product is equal to one.
Besides, the ADBSOM algorithm takes into consideration new sets of constraints.

Finally, the second step of DBSOM and the third step of ADBSOM compute the  partitions on the neurons of the SOM. This is achieved conforming to the optimal solution provided in the paper.

DBSOM and ADBSOM algorithms were evaluated experimentally on two real distributional-valued data sets. ADBSOM outperformed DBSOM on these data sets, especially using toroidal maps. This corroborates the importance of the weighting step of ADBSOM.
We have observed that planar maps, together with the optimized criterion suffer from a pushing-toward-the border effect on the data. This is almost evident in the application on real data. We proposed to overcome such limit of planar maps using a toroidal map that have shown better topology preservation of the map (namely, the topographic error is lower than the one referred to planar maps when all the other map parameters are the same).

From the application on real data, we have observed that the use of adaptive distances reduces generally the topographic error and overcomes the problems related to the scaling of the variables as preprocessing step.
Moreover, ADBSOM algorithms that calculate the relevance weights of the distributional variables for each cluster had the best performance in terms of topological preservation and in terms of compactness and separation of clusters induced by the Voronoi sets associated with the neurons.

Moreover, as a supplementary contribution, we introduced two new Silhouette indexes for SOM, taking together the idea behind the topographic error computation and the clustering structure of the SOM. We have also observed how to compute exactly a Silhouette index when data are described in a Euclidean space in a more efficient way.
% This kind of variables are characterized to take as values one-dimensional probability or frequency distributions on a numeric support.

Finally, the usefulness of DBSOM and ADBSOM algorithms have been  highlighted with their applications to the Human Behavior Recognition and World countries population pyramids distributional-valued data sets.

\section*{Acknowledgments}

The first author would like to thanks
%the anonymous referees for their careful revision, and
to Conselho Nacional de Desenvolvimento Cientifico e Tecnologico - CNPq (303187/2013-1) for its partial financial support.

                \section*{APPENDIX: Fast computation of Silhouette index for Euclidean spaces}
                Let us consider a table $Y$ containing $N$ numerical data vectors $\mathbf{y}_i=[y_{i1},\ldots,y_{iP}]$ of $P$ components. Without loss of generality, the $N$ vectors are clustered in two groups, namely,  A and B, having, respectively, size $N_A$ and $N_B$ such that $N_A+N_B=N$. Let $\bar{y}_{Aj}= n_A^{-1}\sum\limits_{\mathbf{y_i}\in A}y_{ij}$ and $\bar{y}_{Bj}= n_B^{-1}\sum\limits_{\mathbf{y_i}\in B}y_{ij}$, $j=1,\ldots,P$, be the two cluster means, and
                $SSE_{Aj}=\sum\limits_{\mathbf{y_i}\in A}{y_{ij}^2}- n_A (\bar{y}_{Aj})^2$ and $SSE_{Bj}=\sum\limits_{\mathbf{y_i}\in B}{y_{ij}^2}- n_B (\bar{y}_{Bj})^2$ the two sum of squares deviations from the respective cluster means.
                \subsubsection*{The average Euclidean distance between a point to all the other points of a set where it is contained.}
                Let consider that $\mathbf{y}_i\in A$, 
                %and we want to compute 
                the average distance of $\mathbf{y}_i$ with respect all the other vectors in A is computed as follows:
                $$(n_A-1)^{-1}\sum\limits_{\mathbf{y}_k\in A}\sum\limits_{j=1}^P{(y_{ij}-y_{kj})^2}$$
                It is easy to show, for a single variable $j$, that:
                $$\sum\limits_{\mathbf{y}_k\in A}{(y_{ij}-y_{kj})^2}=n_A(y_{ij})^2+\sum\limits_{\mathbf{y}_k\in A}{(y_{kj})^2}-2 y_{ij}\sum\limits_{\mathbf{y}_k\in A}{y_{kj}}=$$
                $$=n_A(y_{ij})^2+\left[SSE_{Aj}+n_A(\bar{y}_{Aj})^2\right]-2n_Ay_{ij}\bar{y}_{Aj}=$$
                $$=n_A\left(y_{ij}-\bar{y}_{Aj}\right)^2+SSE_{Aj}.$$
Then, the average distance is:
%for a single variable $j$, we have that:
$$(n_A-1)^{-1}\sum\limits_{\mathbf{y}_k\in A}
%\sum\limits_{j=1}^P
{(y_{ij}-y_{kj})^2}=\frac{n_A\left(y_{ij}-\bar{y}_{Aj}\right)^2}{n_A-1}+\frac{SSE_{Aj}}{n_A-1}$$
%Then, it follows that $a(i)$ in the Silhouette index is:

                \subsubsection*{The average Euclidean distance of a point from all the other points of a set where it is not contained}
                Let consider that $\mathbf{y}_i\in A$ and we want to compute the average distance of $\mathbf{y}_i$ with respect all the other vectors in B. The average squared Euclidean distance between $\mathbf{y}_i$ and all the other vectors in B, for each variable, is given by:

              $$(n_B)^{-1}\sum\limits_{\mathbf{y}_k\in B}\sum\limits_{j=1}^P{(y_{ij}-y_{kj})^2}$$
                for a single variable $j$, it is easy to show that:
                $$\sum\limits_{\mathbf{y}_k\in B}{(y_{ij}-y_{kj})^2}=n_B(y_{ij})^2+\sum\limits_{\mathbf{y}_k\in B}{(y_{kj})^2}-2 y_{ij}\sum\limits_{\mathbf{y}_k\in B}{y_{kj}}=$$
                $$=n_B(y_{ij})^2+\left[SSE_{Bj}+n_B(\bar{y}_{Bj})^2\right]-2n_By_{ij}\bar{y}_{Bj}=$$
                $$=n_B\left(y_{ij}-\bar{y}_{Bj}\right)^2+SSE_{Bj}.$$
Then, the average distance is that:
%for a single variable $j$, we have that:
$$(n_B)^{-1}\sum\limits_{\mathbf{y}_k\in B}
%\sum\limits_{j=1}^P
{(y_{ij}-y_{kj})^2}=\left(y_{ij}-\bar{y}_{Bj}\right)^2+\frac{SSE_{Bj}}{n_B}$$
        \subsection*{The Silhouette index}
        As stated above, the Silhouette index for a single $\mathbf{y}_i$ is given by:
        $$ s(i)=\frac{b(i)-a(i)}{max[a(i),b(i)]}$$
        where, in the case of two groups A and B, if  $i\in A$:
        $$a(i)=(n_A-1)^{-1}\sum\limits_{\mathbf{y}_k\in A}\sum\limits_{j=1}^P{(y_{ij}-y_{kj})^2}$$
        and
        $$b(i)=(n_B)^{-1}\sum\limits_{\mathbf{y}_k\in B}\sum\limits_{j=1}^P{(y_{ij}-y_{kj})^2}.$$
        If we consider the original formulation, for computing $N$ silhouette indexes the computational complexity is in the order of $\mathcal{O}(N^2)$, if $N$ is sufficiently larger than $P$.
        If we use the formulas suggested here, once computed the averages and the SSE, the computational cost is approximatively of $\mathcal{O}(N)$. In fact, it is possible to compute $a(i)$ and $b(i)$ as follows:
        $$a(i)=\sum\limits_{j=1}^p\left[\frac{n_A\left(y_{ij}-\bar{y}_{Aj}\right)^2}{n_A-1}+\frac{SSE_{Aj}}{n_A-1}\right]$$
        and
        $$b(i)=\sum\limits_{j=1}^p\left[ \left(y_{ij}-\bar{y}_{Bj}\right)^2+\frac{SSE_{Bj}}{n_B}\right].$$
        Considering that the squared L2 Wasserstein distance is an Euclidean distance between quantile functions, that the SSE computed for distribution functions is defined as a sum of squared distance an that the average distributions are Fr\'{e}chet means with respect to the squared $L_2$ Wasserstein distance. The same simplification can be applied for computing the Silhouette Coefficient when data are distributions and they are compared using the squared $L_2$ Wasserstein distance.

        \section*{References}
		
		\bibliography{mybibfile}
		
	\end{document}